\documentclass[conference]{IEEEtran}
\usepackage{epsfig}
\usepackage{endnotes}
\usepackage{times}
\usepackage{balance}
\usepackage{amsfonts}
\usepackage{epsfig}
\usepackage{amssymb}
\usepackage{amscd}
\usepackage{amsmath}
\usepackage{microtype}
\usepackage{array}
\usepackage{epsfig}
\usepackage{multirow}
\usepackage{graphicx}
\usepackage[font={small,bf}]{caption}
\usepackage{listings}
\usepackage{subfigure}
\usepackage{graphicx}
\usepackage[obeyspaces]{url}

\usepackage[
  breaklinks = true,
  unicode = true,
  urlcolor = blue,
  colorlinks = true,
  citecolor = blue,
  linkcolor = red]{hyperref}
\usepackage{tabularx}
\usepackage{epstopdf}
\usepackage{threeparttable}
\usepackage{lipsum}
\usepackage[T1]{fontenc}
\usepackage[utf8]{inputenc}
\usepackage{tcolorbox}
\usepackage{flushend}

\tcbuselibrary{breakable, skins}

\tcbset{%
	label begin/.style={label={#1}},
	label end/.style={after upper=\label{#1}}
}

\newtcolorbox[%
auto counter]{mybox}[2][]{%
	enhanced jigsaw,
	breakable,
	#1}
\usepackage{pifont}
\newcommand{\ignore}[1]{}
\newcommand{\revised}[1]{}

\usepackage{graphicx}
\usepackage{soul}
\usepackage{booktabs}

\newcolumntype{M}[1]{>{\centering\arraybackslash}m{#1}}

\usepackage{tikz}

\newcommand{\hpie}[0]{%
\begin{tikzpicture}
 \draw (0,0) circle (0.7ex);\fill (0,0.7ex) arc (90:270:0.7ex) -- (0,0) -- cycle;
\end{tikzpicture}%
}

\newcommand{\epie}[0]{%
\begin{tikzpicture}
 \draw (0,0) circle (0.7ex);
\end{tikzpicture}%
}

\newcommand{\fpie}[0]{%
\begin{tikzpicture}
 \draw (0,0) circle (0.7ex);\fill (0.7ex,0) arc (0:360:0.7ex) -- (0,0) -- cycle;
\end{tikzpicture}%
}

\usepackage{ctable}
\usepackage{threeparttable}

\newcommand{\BULLET}{\vspace{+.05in} \noindent $\bullet$ \hspace{+.05in}}
\newcommand{\BULLETSMALL}{\vspace{+.01in} \noindent $\bullet$ \hspace{+.05in}}

\newcolumntype{C}[1]{>{\centering\let\newline\\\arraybackslash\hspace{0pt}}m{#1}}
\newcolumntype{L}[1]{>{\raggedright\let\newline\\\arraybackslash\hspace{0pt}}m{#1}}

\makeatletter
\renewcommand\footnoterule{%
  \kern-3\p@
  \hrule\@width.4\columnwidth
  \kern2.6\p@}
  \makeatother

\begin{document}

\title{Understanding IoT Security Through the Data Crystal Ball: Where We Are Now and Where We Are Going to Be}


\author{\IEEEauthorblockN{Nan Zhang$^1$,
Soteris Demetriou$^2$, Xianghang Mi$^1$, Wenrui Diao$^3$, Kan Yuan$^1$, Peiyuan Zong$^4$, Feng Qian$^1$\\ XiaoFeng Wang$^1$, Kai Chen$^4$, Yuan Tian$^5$, Carl A. Gunter$^2$, Kehuan Zhang$^3$, Patrick Tague$^5$ and
Yue-Hsun Lin$^6$}
\IEEEauthorblockA{$^1$\textit{Indiana University, Bloomington} 
\{nz3, xmi, kanyuan, fengqian, xw7\}@indiana.edu}
\IEEEauthorblockA{$^2$\textit{University of Illinois at Urbana-Champaign}
\{sdemetr2, cgunter\}@illinois.edu}
\IEEEauthorblockA{$^3$\textit{The Chinese University of Hong Kong}
\{dw013, khzhang\}@ie.cuhk.edu.hk}
\IEEEauthorblockA{$^4$\textit{Institute of Information Engineering, Chinese Academy of Sciences}
\{zongpeiyuan, chenkai\}@iie.ac.cn}
\IEEEauthorblockA{$^5$\textit{Carnegie Mellon University}
\{yt, tague\}@cmu.edu $^6$\textit{Samsung Research America}
yuehhsun.lin@samsung.com}
}

\maketitle

\begin{abstract}
Inspired by the boom of the consumer IoT market, many device manufacturers, new start-up companies and technology behemoths have jumped into the space. Indeed, in a span of less than 5 years, we have experienced the manifestation of an array of solutions for the smart home, smart cities and even smart cars. Unfortunately, the exciting utility and rapid marketization of IoTs, come at the expense of privacy and security. Online and industry reports, and academic work have revealed a number of attacks on IoT systems, resulting in privacy leakage, property loss and even large-scale availability problems on some of the most influential Internet services (e.g. Netflix, Twitter). To mitigate such threats, a few new solutions have been proposed. However, it is still less clear what are the impacts they can have on the IoT ecosystem. In this work, we aim to perform a comprehensive study on reported attacks and defenses in the realm of IoTs aiming to find out what we know, where the current studies fall short and how to move forward. To this end, we first build a toolkit that searches through massive amount of online data using semantic analysis to identify over 3000 IoT-related articles (papers, reports and news). Further, by clustering such collected data using machine learning technologies, we are able to compare academic views with the findings from industry and other sources, in an attempt to understand the gaps between them, the trend of the IoT security risks and new problems that need further attention. We systemize this process, by proposing a taxonomy for the IoT ecosystem and organizing IoT security into five problem areas. We use this taxonomy as a beacon to assess each IoT work across a number of properties we define. Our assessment reveals that despite the acknowledged and growing concerns on IoT from both industry and academia, relevant security and privacy problems are far from solved. We discuss how each proposed solution can be applied to a problem area and highlight their strengths, assumptions and constraints. We stress the need for a security framework for IoT vendors and discuss the trend of shifting security liability to external or centralized entities. We also identify open research problems and provide suggestions towards a secure IoT ecosystem.

\end{abstract}

\IEEEpeerreviewmaketitle

\setlength{\textfloatsep}{0pt}
\setlength{\intextsep}{0pt}
\setlength{\abovecaptionskip}{2pt}
\setlength{\belowcaptionskip}{0pt}
\setlength{\dbltextfloatsep}{0pt}
\setlength{\dblfloatsep}{0pt}

\section{Introduction}\label{sec:introduction}
The Internet of Things (IoT) is transforming many key industries. Based on a recent estimation, the number of IoT devices deployed worldwide will grow from 5 billion in 2015 to 24 billion in 2020, forming a global market of \$13 trillion dollars~\cite{iot_market}.
One of the key challenges faced by the IoT ecosystem is privacy and security. While this domain shares some problems with conventional wireless sensor networks (WSN), it also has some unique traits. IoT devices are user-centric, are Internet-connected and have more complex software/hardware. As such, potential compromises may cause serious harms (e.g., physical property damage~\cite{url_physical_damage}, bodily injuries~\cite{url_medical_hack}, traffic accidents~\cite{url_tesla_hack}). In addition, remote adversaries become possible due to the connectivity of IoT devices and can readily launch large-scale attacks~\cite{mirai_million}, while the complexities of IoT devices make security vulnerabilities harder to be discovered. The severity and impact of the consequences in tandem with the ease of performing scalable attacks, provide adversaries with strong incentives to launch attacks.

Such IoT systems can be targeted towards either consumers or enterprises following a business-to-consumer (B2C) or business-to-business (B2B) model. However, consumer IoT devices are likely more vulnerable compared to their industrial counterparts that are usually selected and managed by professionals, protected by enterprise firewalls and intrusion detection system (IDS), and their manufacturing is driven by legal and liability contracts. In this work, we focus our attention on IoT devices and systems of which target market consists of end-user consumers. We leave the investigation of security in enterprise IoT systems as future work.

Consumer IoT is an exciting topic not only for academia but also for industry. In fact, individual researchers, industry research labs and cyber-security firms have conducted their own studies. Commonly, these works perform assessments on IoT devices regarding their susceptibility to known attacks and aim to determine whether they suffer from known security issues. Therefore, these works tend to be publicized in white papers, news articles, blog posts rather than appearing in major security academic conferences. Hence, any study on IoT that fails to take into account these works, will provide a very conservative picture of IoT security.

This paper presents the first comprehensive investigation of security for off-the-shelf IoT systems. The overall goal of our study is to provide the research community with a panoramic view of state-of-the-art attacks on modern IoT systems and their defenses, as well as to offer the key knowledge and insights that will evolve the IoT security. Overall, our work systemizes the knowledge of IoT security in the following aspects:

\BULLET Going beyond performing a regular research paper survey, we collect literature from a wide range of sources including academic publications, white papers, news articles, blog posts, etc., to provide a more comprehensive picture of today's IoT security status. Note that collection of such data is by no means trivial, due to the diversity in available information sources, the huge volume of related web documents, and the overlapping of the information from various sources. To address these challenges, we develop a literature retrieval and mining tool that automatically performs web page crawling, semantic extraction, content filtering, and article clustering. The tool generates high-quality clusters each representing a ``story'' of IoT security that is used for subsequent classification and analysis. Overall, our literature database consists of 3,367 white papers, news articles, blog posts, together with 47 academic papers we collect, resulting in 107 unique attack incidents spanning from Q1 of 2010 to Q3 of 2016. All datasets we collect and organized in this study will be made available online~\cite{iot_sok}. The datasets will be updated periodically.


\BULLET After meticulously analyzing the literature from both the academia and industry, we generalize IoT security attacks into five problem areas: \textit{LAN Mistrust}, \textit{Environment Mistrust}, \textit{App Over-privilege}, \textit{No/Weak Authentication} and \textit{Implementation Flaws}. For each problem area, we conduct a detailed investigation on its prevalence, severity, and root causes. We apply the classification to existing defense strategies to align the protection against the threats. Further, we propose new taxonomies for both attacks and defenses based on various properties such as their network communication channel, threat model, stealthiness, and required modifications (for defense). This brings out several insights such as the gap between attacks and defenses and the discrepancy between the academic research findings and industrial practices. 


\BULLET Based on our detailed assessment, we make several key recommendations for improving the state of the art. We identify the need to solidify our understanding of the mechanics of emerging IoT application platforms. We also suggest focusing on threats stemming from the physical environment such as working on voice authentication and gesture control authentication. We identify lack of security solutions for connected cars and suggest isolation of safety critical components from other-purpose components in in-vehicle networks ad architectures. We further reveal an abundance of implementation and authentication flaws in IoT devices and stress the need for academia, industry and legislators to combine forces in designing a security framework for IoT devices. Lastly, we highlight the need for practical, backward compatible, device-independent access control solutions that take into consideration mobile and LAN adversaries in home networks.


\textbf{Paper Organization.} In Section~\ref{sec:background}, we define our scope and the players in the IoT ecosystem; in Section~\ref{sec:systemization}, we present our data collection tool and in Section~\ref{sec:taxonomy} we identify the five main problem areas we discover; in Section~\ref{sec:attack}, we perform an assessment on all collected IoT attacks while in Section~\ref{sec:defense}, we assess the solutions proposed by academia. In Section~\ref{sec:discussion}, we present our insights and recommendations drawn from our assessment. In Section~\ref{sec:related}, we discuss some related work and in Section~\ref{sec:conclusion}, we conclude the paper.

\section{Consumer Internet of Things}\label{sec:background}
In this section, we describe the focus of this work, provide a general picture of the consumer IoT ecosystem, and define the major entities it is comprised of.

\subsection{Scope and Importance}
The business model of IoT can be business-to-consumer (B2C) or business-to-business (B2B). The former, builds products for end-user consumers while the latter targets enterprises. The focus of our study is B2C (or consumer) devices. We believe that end-users are more vulnerable to attacks on IoT since they lack the technical expertise and the resources to deploy in-depth defense solutions that protect them from potential adversaries. On the other hand, enterprises usually enjoy the services of highly trained administrators who can utilize firewalls and intrusion detection systems (IDS) with complex security rules to ensure security and privacy within the corporate network.


Moreover, enterprises usually follow strict protocols in deciding which devices can be plugged-in the corporate network. They are also commonly safeguarded by agreements to protect them from damages incurred by defective devices which exert pressure on IoT vendors in attesting the security of their products. In contrast, companies targeting end-users do not have security as a priority and are mostly driven by the time-to-market. In fact, with the explosion and hype of IoT in recent years, a large number of start-up companies entered the space. Unfortunately, delivering a product first can make the difference between the company's success and demise. Note that there are 357 such companies in home automation. More importantly, the vast majority of them (167 out of 217 that share their data), have less than 10 employees~\cite{hastartup}. Obviously, the priority is to make their products functional and deliver them on time; focusing on security would be an expensive and time-consuming endeavor, due to the need to hire security experts and add security testing into their development cycles.



One may argue that IoT devices are not very different from traditional embedded devices, or WSN devices. However, despite their similarities, IoT devices have some unique characteristics which affect their security. In particular they are (a) user-centric, (b) Internet-connected and, (c) complex. Next, we describe these characteristics.

\BULLETSMALL\textit{User-centric:} IoT devices can play the role of actuators and sensors which can act on or sense the user's physical environment. For example, some IoT devices are equipped with a barometer that allows them to measure air pressure in a room. Some others have motion sensors which enable them to perceive movement in a room. Unlike WSN devices, data from home automation sensors can reveal private user information such as daily activities and presence in a physical space. The former can be utilized by an adversary to profile users while the latter can be utilized in committing theft. On the other hand, actuator devices can assert a change in the physical environment either directly or indirectly. For example, a smart lock can be used to unlock the door of a house whereas a smart thermostat can change the temperature in the room. Again, unlike traditional WSN devices, these devices, once compromised, could inflict serious damages to the user, by allowing a thief to surreptitiously enter a house, or keep the user in perennial hot temperatures which can further result in a huge electricity bill.

\BULLETSMALL\textit{Internet-connected:} IoT devices are connected to the Internet. This connectivity can be direct: devices can gain Internet connection through the cellular network. It can also be indirect: devices can connect to the Internet through an existing connected device (an IoT hub, a home router, or a user smartphone in proximity). This Internet connectivity is what allows the users to remotely control the devices and set-up control rules for them. However, while this enables automation advances and flexibility in control, at the same time it opens new attack surfaces, allowing remote (Internet) adversaries to perform large-scale attacks. For example, it is reported that a botnet based on Mirai took control of millions of Internet-connected home-automation devices with authentication problems~\cite{mirai_million} and then used them in a massive DDoS attack on a domain name server, rendering some of the most popular Internet services unavailable for hours.

\BULLETSMALL\textit{Complex:} Today's IoT devices are complex. Instead of residing on primitive hardware with microcontrollers and KBs of memory (as sensor network motes a decade ago), today’s IoT applications run on powerful hardware with full-blown software stacks. As IoT systems become more sophisticated, they may potentially expose more vulnerabilities that are not easy to discover.


\ignore{
\begin{itemize}
\item \textit{user-centric}; IoT devices are incorporated with many sensors e.g. barometer, contact sensor and/or actuators e.g. lock, heater to perceive user or physical environment and change the physical environment directly (unlock the lock) or indirectly (control air conditioner). Unlike WSN devices, some IoT devices with sensors generate user private data through their everyday activities which when leaked, could be used to picture their behaviors in detail. Others with actuators interact with user and physical environment, which when something goes wrong, could harm physical properties or even cause bodily injuries.

\item \textit{connectivity}; IoT devices are usually connected to the Internet directly though cellular network or indirectly through a hub, a router or a smartphone, which gives them the possibility to be controlled remotely by user. On the other side, connectivity enables more attack surfaces for those vulnerable IoT devices.

\item \textit{interactive}. IoT devices can be set to work together by users, for example unlock the door when smoke alarm is triggered~\cite{nptsdu}. However, managing IoT devices in quantity could be difficult and may result in conflict rules when setting improperly. Another problem is that serious trouble could be raised due to a single point of failure, e.g. a hacked smart coffee machine burns itself without water causes a small fire which triggers a smoke alarm and eventually unlock the door for a thief to enter.

\end{itemize}
}
\begin{figure}[t]
\centering
\includegraphics[width=7cm]{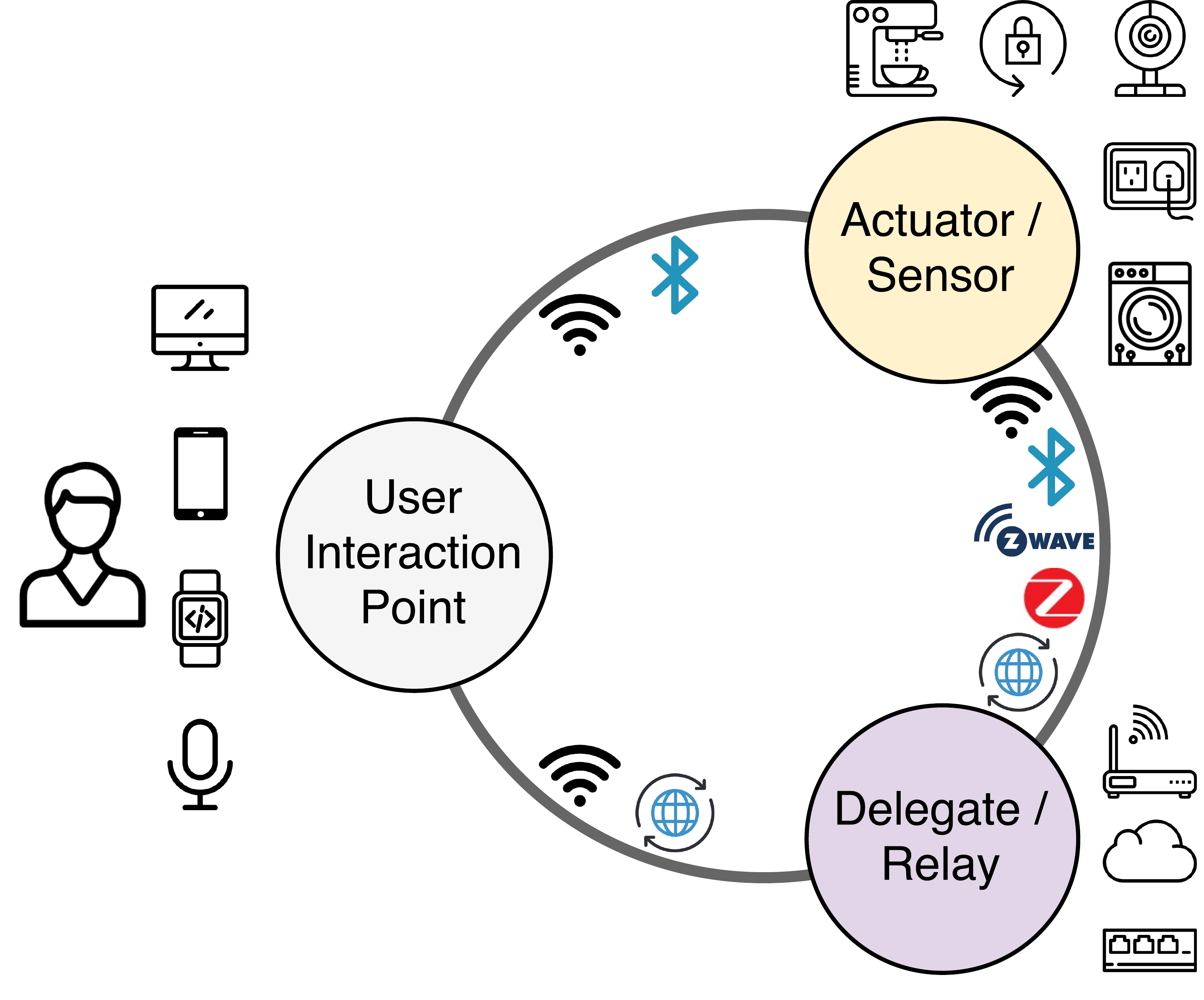}
\caption{Infrastructure of IoT ecosystem.}
\label{fig:ecosystem}
\end{figure}

\subsection{Infrastructure-based categorization}\label{sec:iotcategorization}  Figure~\ref{fig:ecosystem} illustrates the IoT ecosystem. We identify three entities in the ecosystem based on their responsibilities: \textbf{user interaction point (UIP)} which receives user commands and displays information; \textbf{delegate/relay (D/R)} which acts on behalf of users to control IoT devices and/or relay data for users and IoT devices; \textbf{actuator/sensor (A/S)} which changes and/or senses the environment. 

\textit{User interaction point} plays the role of an interface between user and end device. Commonly, UIP entity is assumed by a smartphone. A mobile application developed by an end-device vendor (or a 3rd party) is installed on user smartphones. The user can interact with the mobile app to control the device or get notifications from the device. However, some devices can be controlled with physical buttons while others interface with devices through a dedicated connected device. For example, Amazon Echo accepts voice commands to control other connected devices.

In some architectures, the user-initiated command is mediated by a \textit{delegate/relay} entity. For example, some devices are backed by a cloud service which holds the aggregation logic for various devices. This party is responsible for performing most of the computation. A hub can also take this role, which can work either in conjunction with the cloud service to integrate various devices through diverse communication channels, or perform the computation itself (e.g. Samsung's SmartThings).


After some computation is performed, the final command is delivered to an \textit{actuator}. The actuator changes its state and usually has some effect on the environment. For example, a doorbell can produce sound, a camera can start recording and a coffee machine can start brewing. Not surprisingly, this role is assumed by IoT devices. Note that even though we use a user-initiated command to describe the information flow, these channels are bidirectional. For example, the connected device can be a \textit{sensor} which in effect senses changes in the environment (e.g. motion or temperature). The information is then delivered to a delegate/relay (hub, cloud or both) which computes on the input. The users are notified of the changes through their user interaction point. Things can get more complicated with various integration protocols such as IFTTT (if this then that) which allow the user to set automatic command generation in response to environmental changes. Nevertheless, the interaction can always be reduced to the simple cases we describe here. For example, these automatically-generated commands have to be set up by the user and stored in the delegate/relay. Thus this process can be viewed as an optimization which avoids forcing the user to respond to a command whenever a notification is issued.

\section{Data Collection and Mining}\label{sec:systemization}
Our work goes a step further than a traditional research survey paper. In particular, we utilize data mining techniques to collect literature from a variety of sources to draw a comprehensive picture of IoT security. In this section, we describe our methodology in collecting relevant literature and mining web reports. 

\subsection{High-level approach}
Our goal is to perform a detailed assessment of IoT attacks and proposed solutions. A straightforward approach would be to go through the works published in major security conferences and manually pick those related to IoT security. However, this approach would be incomplete. While most of the solutions are proposed by academia, a large number of real-world attack incidents and IoT device vulnerabilities are revealed by individual researchers and cyber-security firms, which have brought serious IoT vulnerabilities to light. When summarizing and organizing the knowledge in IoT security, we don't want to miss out on these works. In this paper, we complement academic works with literature from a wide range of other sources including white papers, news articles, blog posts, etc. This allows us to draw a comprehensive picture of the state-of-the-art in IoT security (both at the attack and the defense front). To our knowledge, no such SoK exists in the literature.


Collecting these works is a non-trivial endeavor. Most of the attack incidents are published online and manually searching for them does not scale since most of the search engine results are either irrelevant or duplicates. In our work, we utilize Google's search engine APIs~\cite{searchapi} and domain knowledge to construct relevant queries. We further build a smart web crawler which leverages those queries and meaningful combinations of them, to automatically collect anything Google's search engine deems relevant. We then filter out irrelevant reports using a set of heuristics. Next, we cluster reports together utilizing state-of-the-art data mining techniques, to reveal the high-impact reports. This results in 117 clusters with at least 3 web reports, which encompasses 973 web reports in total. Lastly, we manually examine these reports to unearth 76 unique attack instances.

There are many academic papers related to IoT security as well. Since our purpose\ignore{ of this systemization} is to present the current status of IoT research and identify future research areas, we try to include all research works that are IoT security related and cover various parties of the IoT ecosystem and different communication channels. However, we exclude papers that are not IoT-specific, for example, those focusing on finding vulnerabilities or securing communication protocols in general.

In total, we manually collect 47 research papers: 24 of them present offense studies on IoT systems. After manual inspection, we find 31 unique attack instances in these papers which complement the 76 incidents from our web reports for a total of 107 unique attack instances. 30/47 research papers present defenses to counter cyber-attacks on IoT. Note that some papers might describe both attacks and defenses.

\subsection{Collection of Web Reports}
As mentioned before, collecting attack incidents reported by industry and individual researchers is challenging, mainly due to the variety of possible sources. 
To address this challenge, we build a smart web crawler to collect white papers, news articles and blog posts from the World Wide Web. We will refer to these as \textit{web reports} from now on. We first construct relevant queries which our smart crawler automatically executes leveraging Google Search APIs (\textit{query construction and search}). Next, our web crawler invokes a \textit{content parser} module that runs on the collected web reports to eliminate irrelevant results and identify high-impact reports. Next, we describe these functions in detail.


\vspace{5pt}\noindent\textbf{Query Construction and Search}. Our smart web crawler uses the Google search engine APIs to perform an initial collection of candidate web reports. Obviously, the quality of the results depends on the quality of the search queries. One could naively use a large list of IoT-relevant keywords. Following this approach, we end up with very noisy results: the majority of the articles are not relevant to IoT security. In our work, we follow a more systematic approach in constructing such queries. In particular, we have constructed three keyword sets: (a) \textit{general IoT keywords} set; (b) \textit{brand devices} set; (c) \textit{security keywords} set. The \textit{general IoT keywords} set includes keywords describing high-level categories of IoT devices, such as ``wearable'', ``home automation'', etc. The \textit{brand devices} set includes the \textit{brand} and \textit{name} of an IoT device. We extract these from two IoT product collection platforms (i.e. \url{iotlist.co}~\cite{iotlist} and \url{smarthomedb.com}~\cite{smarthomedb}. Lastly, the \textit{security keywords} set includes words related to offensive security such as ``hack'', ``vulnerability'', ``malicious'', etc. We also include variations of these words. For example, a noun can appear in its singular and plural form (attacker, attacker-\textit{s}) while a verb can appear in its infinitive form, past tense and present participle (e.g. attack, attack-\textit{ed}, attack-\textit{ing}). Our queries are automatically constructed by the crawler which picks combinations of words from these 3 sets: For every word in set (a), a unique combination of 4 words from set (c) is picked. The process is repeated for every word of the (b) set. This approach results in 2000 queries. For every query, the crawler picks the first 100 results. We end up with 100,000 unique URLs.

During this data collection, we observe that web articles and news reports are usually published within a few days of the release of the original incident (usually published through a research paper, white paper, technical report or blog post). However, the search engine results, tend to rank more recent web reports higher. This leaves out older works which could in principle be equally as important. To tackle this, we introduce a time constraint to our queries: every query is repeated with a time constraint spanning from the first week of the first quarter of 2010 to the last week of the 3rd quarter of 2016. This results in 17,000 more unique URLs for a total of 117,359 unique URLs. All collected web reports are input into a database which will be made available to the community~\cite{iot_sok} and will be updated periodically.




\begin{figure}[t]
\centering
\includegraphics[width=8cm]{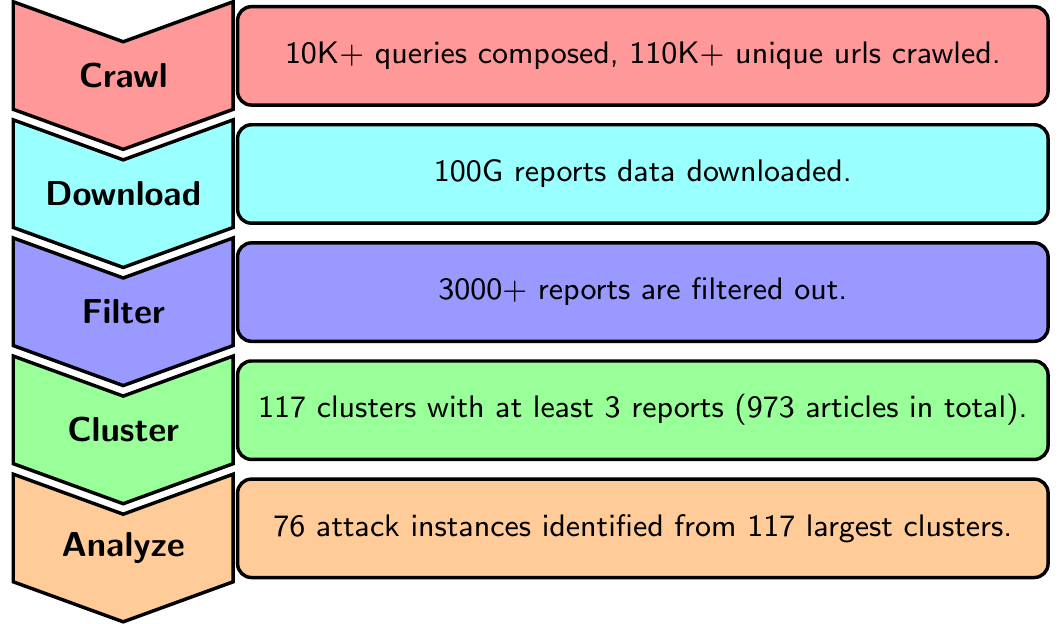}
\caption{Workflow -- Collection of web reports.}
\label{fig:collection}
\end{figure}

\vspace{5pt}\noindent\textbf{Content Parser}. Given the huge amount of web reports collected by our smart crawler, finding attack incidents and their original published source could be very challenging. To address this, we build a content parser module which is composed of two components: (1) a content filter and, (2) an article clustering as illustrated in Figure~\ref{fig:collection}.


\vspace{5pt}\noindent\textit{1) Content Filter}. The main purpose of the content filter is to quickly eliminate the web reports that are not relevant to consumer IoT security. The filter proceeds in two consecutive phases. During the first phase, the content filter eliminates sites that are not web reports at all. During the second phase, it eliminates web reports that are not relevant.

Upon manual inspection of the collected data, we observe that web reports with no hyperlink are usually neither web articles nor blog posts. For example, we find cases where a collected report is merely a collection of keywords. Moreover, sites with too many hyperlinks are either sitemaps or spam. Finally, sites with too many dates tend to be report aggregation sites. We use these observations in the first phase to eliminate candidate reports which have no hyperlinks, have hyperlinks which amount to more than half of the text, or have more than 30 dates. The thresholds are set following a trial and error approach. Next, the survived articles are further checked to ensure that all keywords (or variations of them) from the original query indeed appear in HTML source code. This process results in a reduction from 117,359 candidate web reports to 15,285. We manually sample 200 of the eliminated URLs and inspect them. We find that 100\% of them are true negatives. Note that our goal in this phase is to reduce false positives without eliminating interesting URLs. Evidently, our approach is effective in achieving that.



The second phase of content filtering aims to perform a more in-depth filtering. In particular, it focuses on the actual text content of the candidate report to ensure its relevancy. At this step, the parser utilizes diffbot~\cite{diffbot} to extract the \textit{title}, \textit{content}, \textit{author} and \textit{publish date}~\footnote{\textit{diffbot} can achieve 96.8\% precision and 97.8\% recall.}. Next, the tool checks again for the presence of the original query keywords, but specifically in the \textit{title} and \textit{content} of the report. This is needed to eliminate cases where keywords appear in other places in the source code, such as JavaScript or ads code. The second filtering phase results in 3367 unique URLs. To validate the relevance of the survived web reports, we randomly pick 300 of these reports and check them. We find that 95\% of them are indeed IoT security related articles. We conjecture that the 5\% error stems from the extraction error of diffbot. For example, diffbot may fail to exclude comments from the content. Thus a keyword (or a combination of keywords) might appear in such extraneous content which our tool will match.


\vspace{5pt}\noindent\textit{2) Article Clustering}. Once noise is eliminated from the collected web reports, we focus on identifying high-impact reports. To achieve this, we use clustering to aggregate web articles reporting similar attack incidents.

We first process the title and content of each report to derive a set of words. In order to capture the importance of the title over the content, we assign it a \textit{Title Weight (TW)}. Then we leverage standard natural language processing techniques such as tokenization, removal of stopwords, stemming and punctuation, number filtering, to derive a set of weighted words from the title and content. Next, we extract features vectors utilizing a keyword extraction algorithm tf-idf~\cite{tfidf}. Finally, we construct a distance matrix using cosine similarity of the extracted feature vectors and generate a minimum spanning tree. We prune the edges that don't satisfy the \textit{Cluster Threshold (CT)} and get a forest wherein each tree is a cluster.

Selecting the optimal values for the \textit{Title Weight} and the \textit{Cluster Threshold} is not straightforward. To address this, we perform an evaluation of different combinations of TW and CT values. We construct the ground truth by sampling 50 articles that cover 15 different attack topics and manually label them into different clusters. Then we try different combinations of values to automatically cluster the 50 articles together with the rest reports. The effectiveness of different combinations is measured by the level to which those labeled articles are correctly clustered together and separated in different clusters. Based on this, we select the optimal values for \textit{TW} and \textit{CT}.

\section{IoT Security Problem Areas}\label{sec:taxonomy}

Our in-depth study of web reports and academic papers resulted in a collection of IoT security issues. We identify five major problem areas we find in the space, which form the pillars of our detailed assessment described in subsequent sections.

\vspace{3pt}\noindent\textbf{LAN Mistrust}. A LAN Mistrust problem occurs when the IoT device trusts the local network that is connected to (i.e., local WiFi). In these scenarios, the IoT device does not take any step into authenticating commands and requests. For example, in a home area network, some devices rely on the network-level authentication (e.g. WPA2-PSK), assuming that once a device can authenticate to the network then it should be trusted. However, this leaves them vulnerable to attacks from 3rd party programs running on authenticated devices (e.g. smartphones, laptops etc.).


\vspace{3pt}\noindent\textbf{Environment Mistrust}. An environment mistrust problem occurs when the IoT device trusts the physical environment within which it is placed. Erroneous environment trust might lead to physical attacks:  an adversary can move the device (point a camera into a different direction). Moreover, this problem area encompasses threats stemming from a proximity attacker who can use signals propagated at physical medium (e.g. gestures or acoustic signals) to command the user interaction point into executing an action.


\vspace{3pt}\noindent\textbf{App Over-privilege}. A number of roles in the IoT ecosystem can carry third-party apps. For example, smartphones allow the installation of third-party apps which can control one or more IoT devices; integration platforms such as Samsung SmartThings allow developers to submit and users to install smart apps to control their IoT devices. However, sometimes these apps can manipulate the coarse granularity of their systems' authorization models to access devices and data they are not supposed to. For example, any app on Android can leverage side channels involving inertial measurement unit (IMU) sensors to profile the device or the user; an app on Android with Bluetooth permission can access any Bluetooth device in the vicinity; an app with a subset of privileges on an IoT device can perform unauthorized actions. Note that we treat most of the smartphone app side-channel attacks as possible due to over-privilege: an app should not be able to access channels which reveal more information than the app's needs for its normal operations.

\vspace{3pt}\noindent\textbf{No/Weak Authentication}. If a work is focused on exploiting or addressing no/weak authentication issues, then it is applicable to this problem area. For example, Bluetooth devices may use the default \texttt{0000} or \texttt{1234} passwords during pairing. User passwords could be subjected to dictionary attacks or can be brute-forced within realistic time bounds. In general, if a target entity does not use authentication at all then it is also applicable. An instance of this problem could be an IoT device which does not authenticate remote commands or firmware updates. If an IoT device authenticates remote requests but does not authenticate LAN requests, then this is reduced to a LAN Mistrust problem.



\vspace{3pt}\noindent\textbf{Implementation Flaws}. This problem area aims to capture attacks and defenses targeting implementation flaws. Such implementation flaws could include (a) leakage of hardcoded credentials, (b) cross-site scripting (XSS), (c) open ports and enabled debugging functionality and, (d) transferring sensitive information over the network in plaintext.


\vspace{3pt}
As discussed above, some offensive works might have multiple attack incidents that cover several problem areas. In addition, some defense works could solve multiple problems at the same time. During our analysis, we evaluate each attack incident differently and discuss each defense solution in the context of all applicable problem areas.


\section{Security Hazards of IoT Systems}\label{sec:attack}

To systematically assess the current offensive works of IoT security conducted by both industry and academia, we propose and define 9 properties across which the works will be compared. We further provide statistics to reveal trends and assess the attack incidents collected in web reports and academic papers.

\subsection{Assessment Properties for Attacks} \label{sec:attackmetrics} Our work performs an in-depth assessment of attacks on the IoT ecosystem, which are reported in web reports and academic papers. Our assessment utilizes some fundamental pillars of evaluation (which we call properties) to systematically study those works. Before elaborating on the results of the assessment, we define the assessment properties.

\vspace {5pt}\noindent\textbf{Target Entity}. As mentioned in Section~\ref{sec:iotcategorization}, the IoT ecosystem consists of three main entities: (a) actuator/sensor (A/S); (b) delegate/relay (D/R) and; (c) user interface point (UIP). The A/S is typically the \textit{IoT device (A/S)}. The D/R could be a \textit{hub or router (D/R)}, or a \textit{cloud server (D/R)}. Lastly, the UIP is commonly the user's \textit{smartphone (UIP)} or a dedicated interface \textit{IoT device (UIP)}.

\vspace {5pt}\noindent\textbf{Device Type}. IoT devices can have diverse features and functionalities. With this property, we aim to differentiate among different types of devices. In particular, we classify each attacked device in one of the following categories: (a) \textit{Home automation} devices; (b) \textit{Wearable} devices (including wearable medical devices); (c) \textit{Hubs and routers}; (d) \textit{Smartphones} and; (e) \textit{Vehicles} (including aftermarket IoT accessories such as OBD (On-board diagnostics) dongles).

\begin{figure}[tb]	
	\centering
	\subfigure[Web reports distribution.]{\includegraphics[width=0.48\columnwidth]{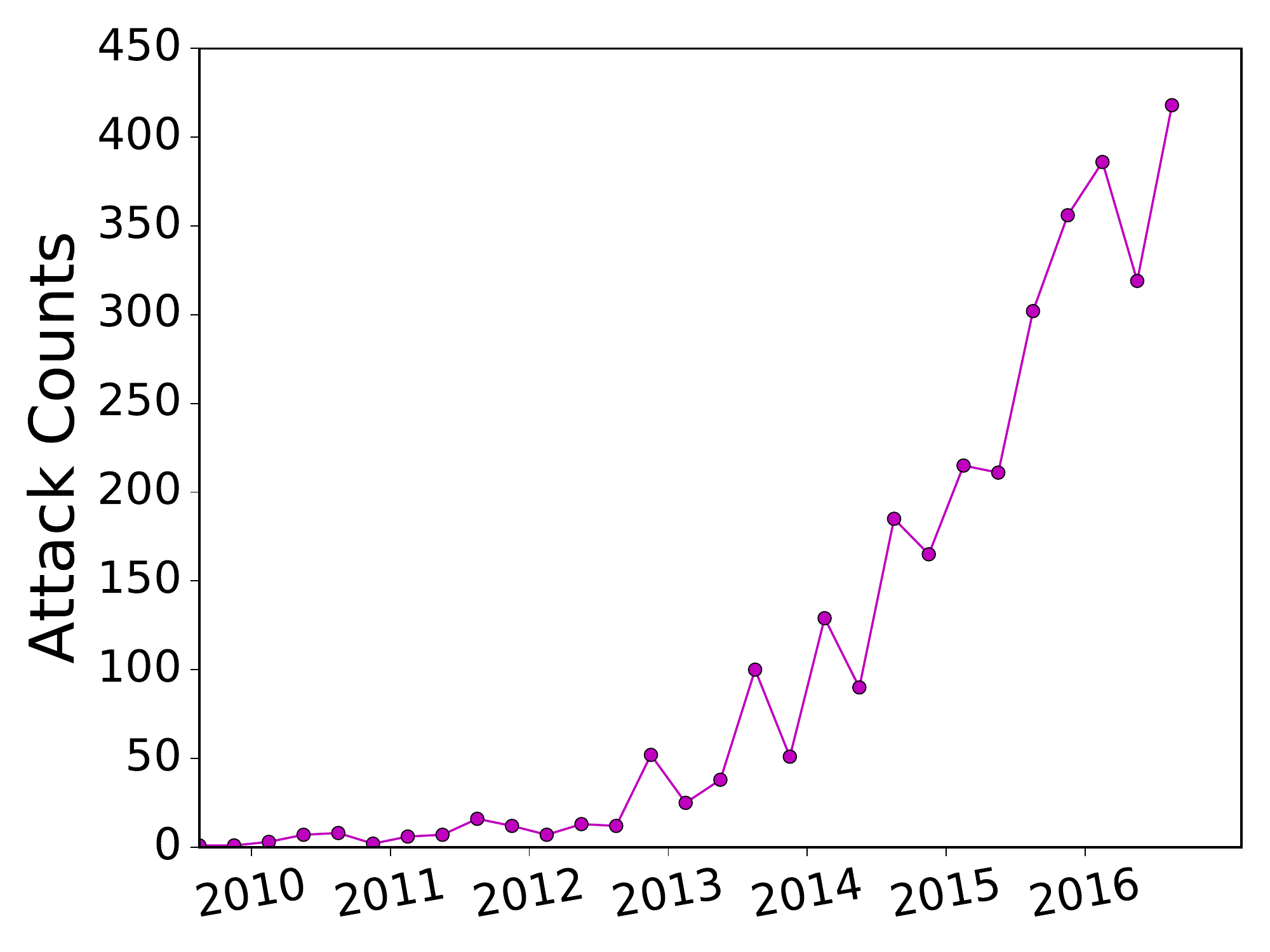}\label{fig:trendbyquater}}
	\quad
	\subfigure[Attack incidents distribution.]{\includegraphics[width=0.467\columnwidth]{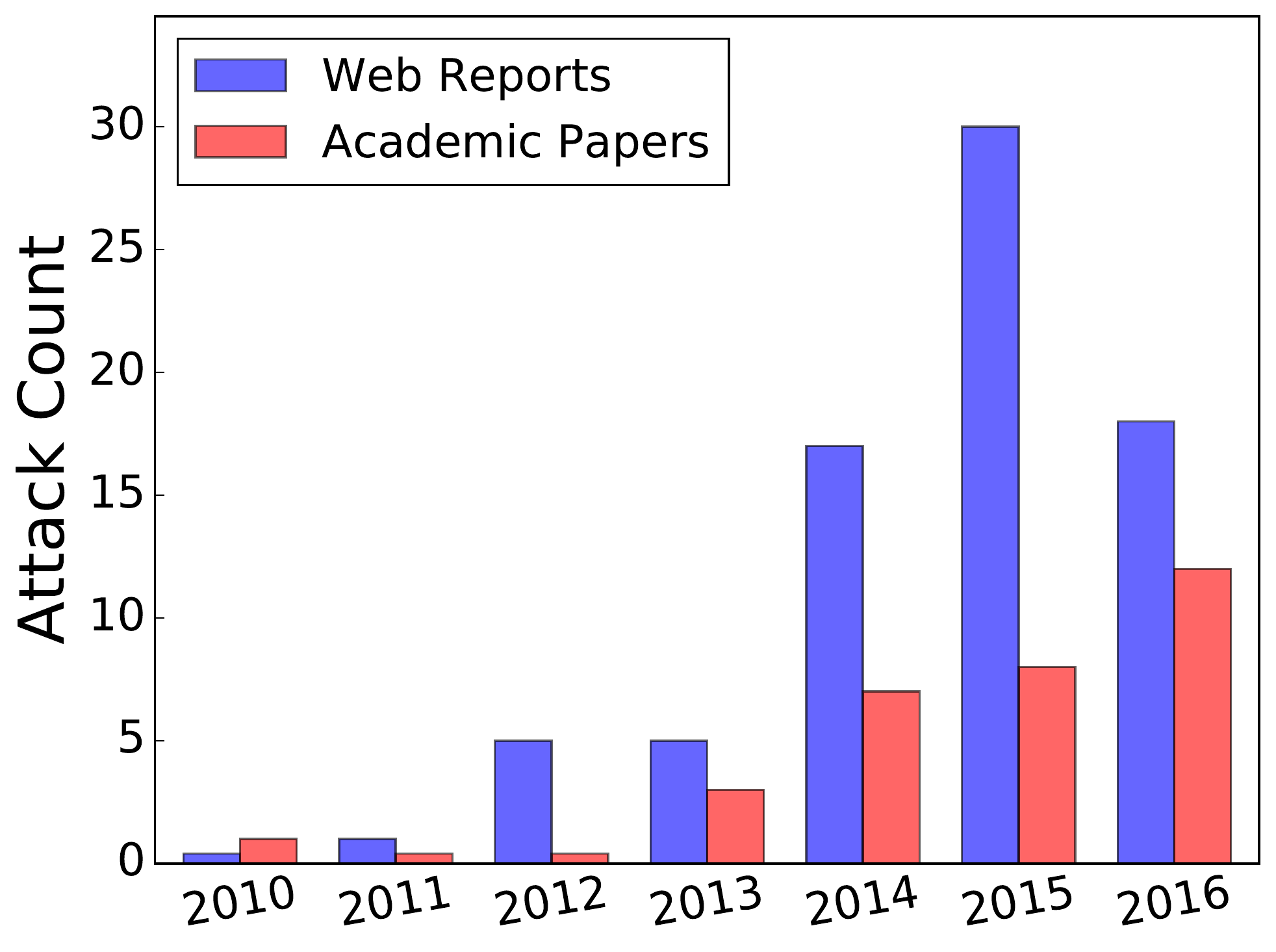}\label{fig:trendbyyear}}
	\caption{Number of published vulnerabilities from 2010 to 2016.}
\end{figure}

\begin{figure*}[!t]
  \footnotesize
  \centering
  \subfigure[Direct consequences per year.]{\includegraphics[width=0.487\columnwidth]{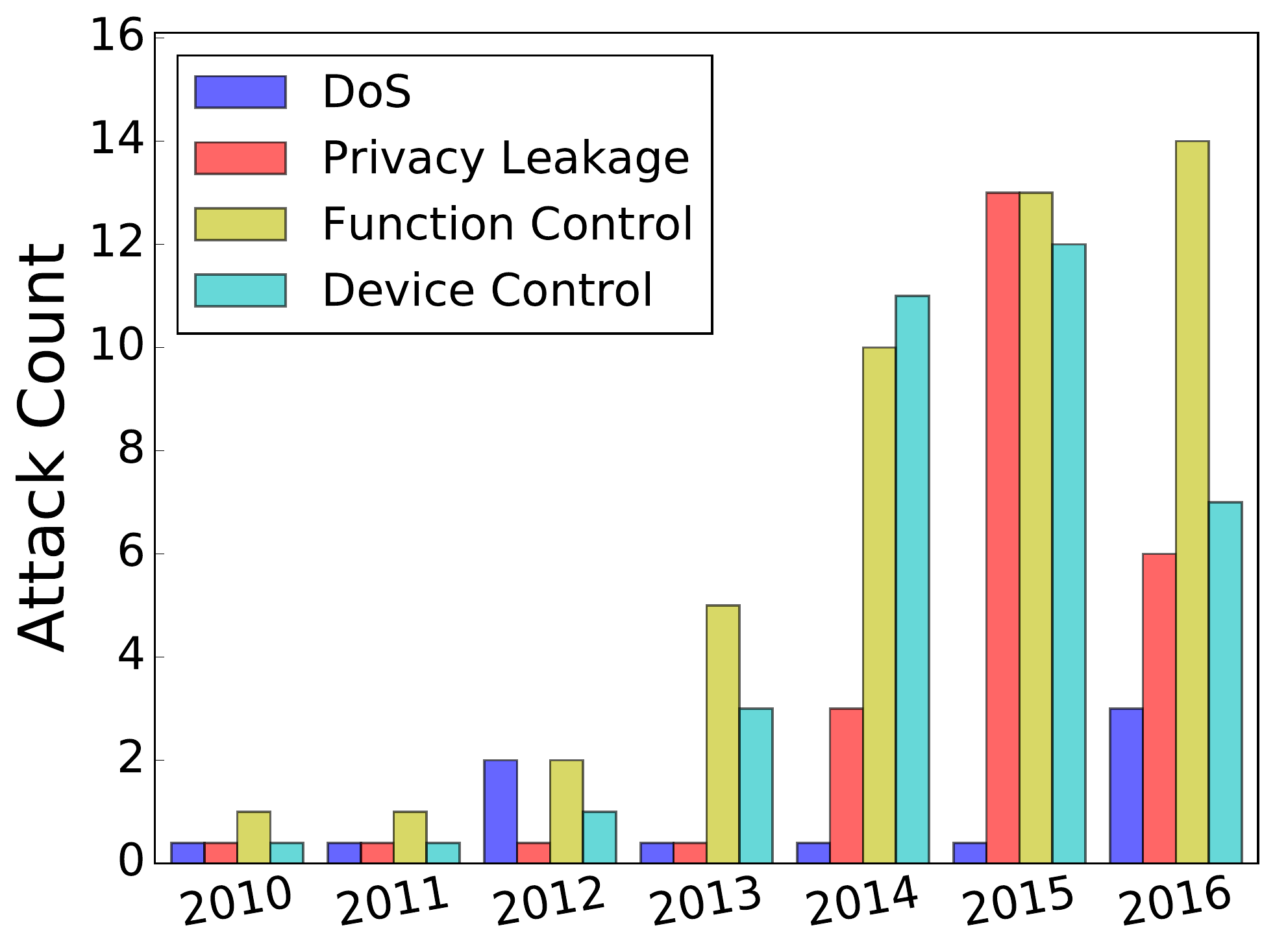}\label{fig:year_direct_overall}}
  \hfill
  \subfigure[Direct consequences per entity.]{\includegraphics[width=0.478\columnwidth]{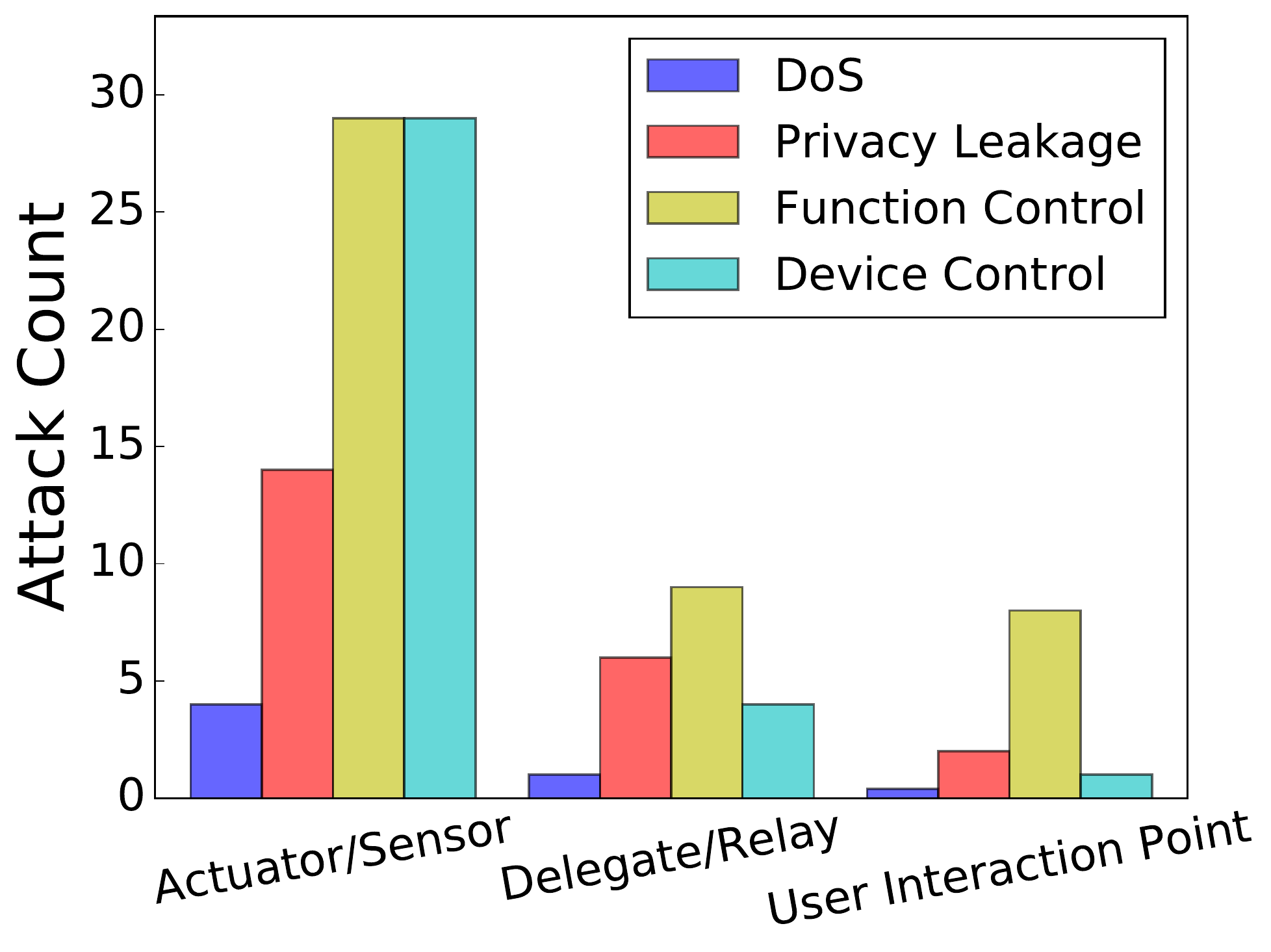}\label{fig:role_direct_overall}}
  \hfill
  \subfigure[Device type per year.]{\includegraphics[width=0.48\columnwidth]{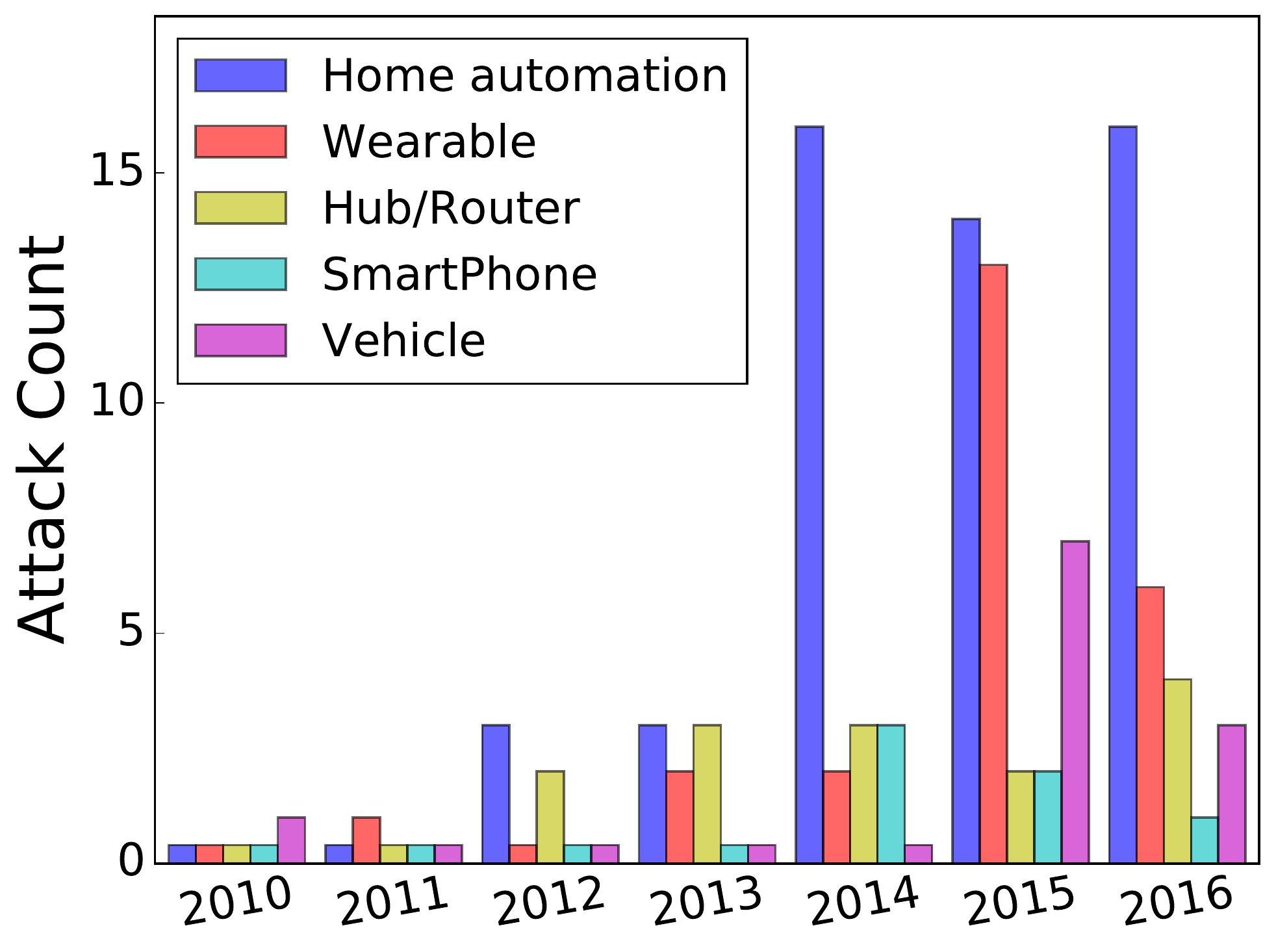}\label{fig:year_deviceType_overall}}
  \hfill
  \subfigure[Affected communication channels.]{\includegraphics[width=0.48\columnwidth]{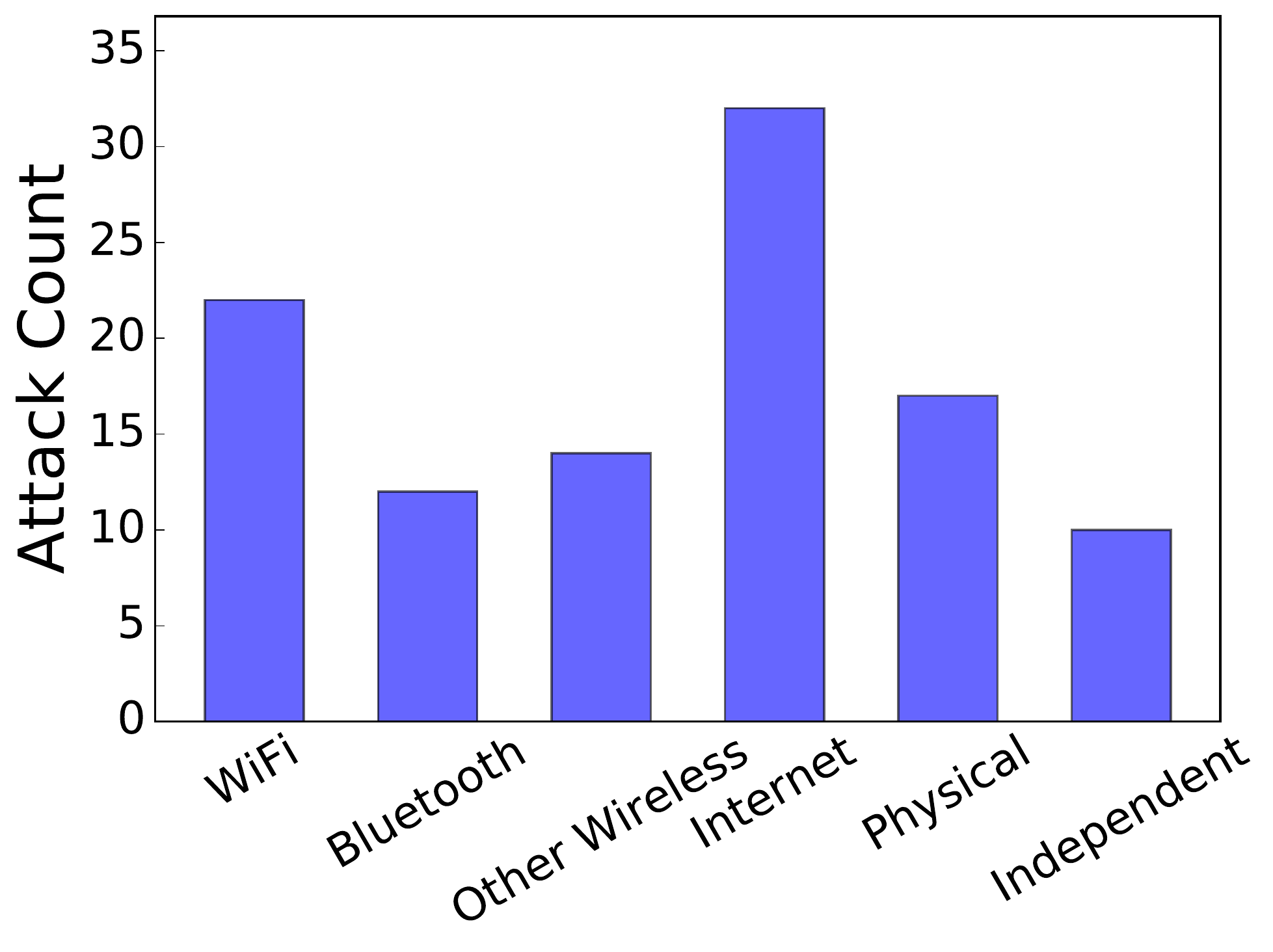}\label{fig:bar_channel_overall}}
  \caption{Trend of attacks mined from 107 unique IoT attack incidents.}
\end{figure*}


\vspace {5pt}\noindent\textbf{Threat Model}. As usual, the threat model describes the capabilities of adversaries. We categorize such adversaries on IoT into five general categories. (a) \textit{Malicious app} describes an adversary who can install a malicious app on a user's smartphone, an IoT hub, an IoT device or an IoT backend cloud server. (b) A \textit{LAN} adversary is authenticated in the LAN environment where the IoT devices are connected. Such an adversary could be a guest device in the home network. (c) A \textit{proximity} adversary can take advantage of their spatial proximity to an IoT device to attack the device. An example of a proximity attacker could be one who sniffs BLE advertisements. (d) A \textit{remote} attacker can launch a remote (through the Internet) attack on an IoT device or other parties. Lastly, (e) a \textit{physical} adversary is physically near the IoT device. For example, a thief might rotate a security camera to make its field of view not cover the point of interest.

\vspace {5pt}\noindent\textbf{Communication Channel}. The communication channels encompass the different ways information can be exchanged among the user or the user interaction point (UIP) and the delegate/relay (D/R) or the IoT device (A/S) itself. We identify five major communication channels: (a) \textit{WiFi}; (b) \textit{Bluetooth}; (c) \textit{other wireless} protocols such as Zigbee, Z-Wave, Thread, X10 Powerline; (d) the \textit{Internet} channel for remote connections and ; (e) the \textit{physical medium} which describes interactions that are mostly sensed, in contrast with traditional network protocols. Such channels can be acoustic signals, gestures and physical touch.

\vspace {5pt}\noindent\textbf{Direct Consequences}. When an attack successfully launches, it will incur (a) denial of service (\textit{DoS}), if the service this device provides is unavailable to users; (b) \textit{privacy leakage}, if the attack can acquire sensitive information of this device; (c) \textit{function control}, if the attacker can take control some or all of the functions this device provides to users; (d) \textit{device control}, if the attacker can gain some privileges or capabilities this device does not provide to users, for example running arbitrary commands in shell.

\vspace {5pt}\noindent\textbf{Technique Scalability}. An attack might be applicable to (a) \textit{one} device; (b) \textit{some devices}; or (c) \textit{any device}. For example, if an attack requires physical access to a particular device then we deem it to affect only one device at a time from a scalability perspective. However, if the attack can affect multiple devices at once (e.g. all devices connected to the LAN network) then it is of the second category. Lastly, if the attack can be performed remotely, at an Internet-wide scale, and every single device in the system is affected then we assign the \textit{any device} value to the scalability property.

\vspace {5pt}\noindent\textbf{Stealthiness}. This property aims to capture whether an attack can be unnoticeable to the user. In particular, an attack can (a) be \textit{unnoticeable}, if the user can be completely unaware that the attack has happened; \ignore{(b) have some \textit{display interference}---when the attack happens the user can notice an alarm/notification on her smartphone, or IoT device display; (c) require \textit{user interaction}---when the attack happens the user needs to confirm the potential device status change and;} (b) result in an \textit{environment change}---the attack results in a change in the environment (e.g. door opens, temperature increases, bell rings, vehicle stops, etc.).

\vspace {5pt}\noindent\textbf{Responsible Disclosure}. We also keep track of whether the attacks have been reported to the affected vendors before their publication. Responsible disclosure is paramount to provide the vendor the opportunity to fix the problem before it becomes public knowledge (and as such accessible by adversaries). We collect such information from academic papers and online articles which report the attack. If there is no mention of vulnerability disclosure, we mark it as \textit{not specified}.


\vspace {5pt}\noindent\textbf{Vulnerability Fix}. When we examine responsible disclosure, we also take notes on how vendors respond to this vulnerability and whether the vendor did or will release a fix. This property helps us identify which attacks the vendors deem as important and whether it is feasible for them to fix the reported issues. If there is no mention of vendor response or whether the vulnerability is fixed, we mark it as \textit{not specified}.

\subsection{Trends of IoT Attack Incidents}

Figure~\ref{fig:trendbyquater} plots the distribution of the 3367 collected IoT security web reports, spanning from the first quarter (Q1) of 2010 until the third quarter (Q3) of 2016. Evidently, there is a nearly exponential increase in web reports on an annual basis reported in the WWW. This insinuates an increasing interest in the industry and the media in IoT security.


After manually scrutinizing the contents of the web reports in the 117 largest clusters we collect, we find 76 unique attack incidents. Most of them are technical reports or blog posts published by cyber-security companies and individual researchers. These unique attack incidents, together with 31 unique attacks reported by 24 academic papers, are assessed and discussed in detail in Section~\ref{sec:attackdiscussion}. Figure~\ref{fig:trendbyyear} shows the frequency distribution across time of the 76 attacks in web reports and the 31 attacks revealed in academic papers.  We do observe that while there is an increase in web reports in 2016, the unique incidents they report actually decrease from 2015. This is because some web reports discuss incidents that happen earlier in time. In addition, our analysis does not include Q4 of 2016. However, the takeaway here is that \textit{there is a significant increase in attacks reported from both industry and academia since 2010.}

\begin{table*}[!t]
\caption{Assessment of Attacks on the Internet of Things}
\label{tab:attackMetrics}
\scriptsize

\begin{center}
  \def\arraystretch{1.5}
  \setlength\tabcolsep{0.9pt} 
  \begin{tabular}{ l  r | r  r  r | r  r  r | r  r  r | r  r  r | r  r  r}
  \specialrule{.12em}{.05em}{.05em}
  \multicolumn{2}{l|}{}& \multicolumn{3}{c|}{\textbf{LAN Mistrust}} 
  		& \multicolumn{3}{c|}{\textbf{Environment Mistrust}} 
  		& \multicolumn{3}{c|}{\textbf{App Over-privilege}} 
  		& \multicolumn{3}{c|}{\textbf{No/Weak Authentication}} 
  		& \multicolumn{3}{c}{\textbf{Implementation Flaw}} \\
	
  \specialrule{.08em}{.05em}{.6em}
  
  \multicolumn{2}{l|}{\shortstack[l]{\textbf{R = Report count;} \\ \textbf{P = Paper count; Ref = References}}} & \textbf{R} & \textbf{P} & \textbf{Ref} & \textbf{R} & \textbf{P} & \textbf{Ref} & \textbf{R} & \textbf{P} & \textbf{Ref} & \textbf{R} & \textbf{P} & \textbf{Ref} & \textbf{R} & \textbf{P} & \textbf{Ref} \\ \specialrule{.08em}{.05em}{.05em}

  \multicolumn{2}{l|}{\textbf{Total}} 
  		& \textbf{3} & \textbf{2} & \cite{a6}\cite{d28} 
  		& \textbf{5} & \textbf{5} & \cite{d29}\cite{a9}\cite{d32}\cite{a23}\cite{a24} 
  		& \textbf{0} & \textbf{8} & \cite{a2}\cite{d33}\cite{a10}\cite{a12}\cite{d8}\cite{d27}\cite{a18}\cite{a22} 
  		& \textbf{17} & \textbf{4} & \cite{d29}\cite{a16}\cite{a17}\cite{d34} 
  		& \textbf{51} & \textbf{12} & \cite{a1}\cite{a3}\cite{a4}\cite{a11}\cite{a17}\cite{a20}\cite{a21} 
  		\\ \specialrule{.08em}{.05em}{.05em}
  
  
  \multirow{5}{*}{\textbf{Target Entities}} & IoT device (A/S) 
  		& \textbf{2}& \textbf{1} & \cite{d28} 
  		& \textbf{2}& \textbf{1} & \cite{d29} 
  		& \textbf{0}& \textbf{5} & \cite{d33}\cite{a10}\cite{a12}\cite{a18}\cite{a22} 
  		& \textbf{12}& \textbf{4} & \cite{d29}\cite{a16}\cite{a17}\cite{d34} 
  		& \textbf{40}& \textbf{9} & \cite{a1}\cite{a4}\cite{a11}\cite{a17}\cite{a20}\cite{a21} 
  		\\ 
  
  & Hub and router (D/R)
  		& \textbf{1}& \textbf{1} & \cite{a6} 
  		& \textbf{1}& \textbf{0} & 
  		& \textbf{0}& \textbf{0} & 
  		& \textbf{3}& \textbf{0} & 
  		& \textbf{5}& \textbf{3} & \cite{a1}\cite{a3} 
  		\\ 
  
  & Cloud server (D/R)
   		& \textbf{0}& \textbf{0} & 
  		& \textbf{0}& \textbf{0} & 
  		& \textbf{0}& \textbf{1} & \cite{a2} 
  		& \textbf{0}& \textbf{0} & 
  		& \textbf{5}& \textbf{0} & 
  		\\
  
  & Smartphone (UIP)
  		& \textbf{0}& \textbf{0} & 
  		& \textbf{1}& \textbf{3} & \cite{a9}\cite{a23}\cite{a24} 
  		& \textbf{0}& \textbf{2} & \cite{d8}\cite{d27} 
  		& \textbf{1}& \textbf{0} & 
  		& \textbf{0}& \textbf{0} & 
  		\\ 
  
  & IoT device (UIP)
  		& \textbf{0}& \textbf{0} & 
  		& \textbf{1}& \textbf{1} & \cite{d32} 
  		& \textbf{0}& \textbf{0} & 
  		& \textbf{1}& \textbf{0} & 
  		& \textbf{1}& \textbf{0} & 
  		\\ \hline

  
  \multirow{5}{*}{\shortstack[l]{\textbf{Device} \\ \textbf{Type}}} & Home automation
  		& \textbf{1}& \textbf{1} & \cite{d28} 
  		& \textbf{3}& \textbf{1} & \cite{d29} 
  		& \textbf{0}& \textbf{1} & \cite{a2} 
  		& \textbf{5}& \textbf{1} & \cite{d29} 
  		& \textbf{35}& \textbf{4} & \cite{a1}\cite{a4} 
  		\\ 
  
  & Wearable
  		& \textbf{1}& \textbf{0} & 
  		& \textbf{0}& \textbf{1} & \cite{d32} 
  		& \textbf{0}& \textbf{5} & \cite{d33}\cite{a10}\cite{a12}\cite{a18}\cite{a22} 
  		& \textbf{7}& \textbf{1} & \cite{d34} 
  		& \textbf{8}& \textbf{1} & \cite{a11} 
  		\\ 
  
  & Hub and router
  		& \textbf{1}& \textbf{1} & \cite{a6} 
  		& \textbf{1}& \textbf{0} & 
  		& \textbf{0}& \textbf{0} & 
  		& \textbf{3}& \textbf{0} & 
  		& \textbf{5}& \textbf{3} & \cite{a1}\cite{a3} 
  		\\ 
  
  & Smartphone 
  		& \textbf{0}& \textbf{0} & 
  		& \textbf{1}& \textbf{3} & \cite{a9}\cite{a23}\cite{a24} 
  		& \textbf{0}& \textbf{2} & \cite{d8}\cite{d27}
  		& \textbf{0}& \textbf{0} & 
  		& \textbf{0}& \textbf{0} & 
  		\\ 
  
  & Vehicle
  		& \textbf{0}& \textbf{0} & 
  		& \textbf{0}& \textbf{0} & 
  		& \textbf{0}& \textbf{0} & 
  		& \textbf{2}& \textbf{2} & \cite{a16}\cite{a17} 
  		& \textbf{3}& \textbf{4} & \cite{a17}\cite{a20}\cite{a21} 
  		\\ \hline

  
  \multirow{5}{*}{\textbf{Threat Model}} & Malicious app
  		& \textbf{0}& \textbf{0} & 
  		& \textbf{0}& \textbf{2} & \cite{a23}\cite{a24}
  		& \textbf{0}& \textbf{8} & \cite{a2}\cite{d33}\cite{a10}\cite{a12}\cite{d8}\cite{d27}\cite{a18}\cite{a22} 
  		& \textbf{1}& \textbf{0} & 
  		& \textbf{3}& \textbf{0} & 
  		\\ 
  
  & LAN
  		& \textbf{3}& \textbf{2} & \cite{a6}\cite{d28} 
  		& \textbf{0}& \textbf{0} & 
  		& \textbf{0}& \textbf{0} & 
  		& \textbf{2}& \textbf{0} & 
  		& \textbf{12}& \textbf{2} & \cite{a4} 
  		\\ 
  
  & Proximity
  		& \textbf{0}& \textbf{0} & 
  		& \textbf{4}& \textbf{3} & \cite{d29}\cite{a9}\cite{d32} 
  		& \textbf{0}& \textbf{0} & 
  		& \textbf{7}& \textbf{3} & \cite{d29}\cite{a17}\cite{d34}
  		& \textbf{13}& \textbf{2} & \cite{a11}\cite{a17} 
  		\\ 
  
  & Remote
  		& \textbf{0}& \textbf{0} & 
  		& \textbf{0}& \textbf{0} & 
  		& \textbf{0}& \textbf{0} & 
  		& \textbf{6}& \textbf{0} & 
  		& \textbf{18}& \textbf{4} & \cite{a3}\cite{a4}\cite{a20} 
  		\\ 
  
  & Physical
  		& \textbf{0}& \textbf{0} & 
  		& \textbf{1}& \textbf{0} & 
  		& \textbf{0}& \textbf{0} & 
  		& \textbf{1}& \textbf{1} & \cite{a16} 
  		& \textbf{5}& \textbf{4} & \cite{a1}\cite{a20}\cite{a21} 
  		\\ \hline

  
  \multirow{6}{*}{\shortstack[l]{\textbf{Communication} \\ \textbf{Channel}}} & WiFi
  		& \textbf{3}& \textbf{2} & \cite{a6}\cite{d28}
  		& \textbf{0}& \textbf{0} & 
  		& \textbf{0}& \textbf{0} & 
  		& \textbf{1}& \textbf{0} & 
  		& \textbf{15}& \textbf{1} & \cite{a4}
  		\\ 
  
  & Bluetooth
  		& \textbf{0}& \textbf{0} & 
  		& \textbf{0}& \textbf{1} & \cite{d29} 
  		& \textbf{0}& \textbf{1} & \cite{d33} 
  		& \textbf{5}& \textbf{2} & \cite{d29}\cite{a17} 
  		& \textbf{1}& \textbf{2} & \cite{a11}\cite{a17} 
  		\\ 
  
  & Other wireless
  		& \textbf{0}& \textbf{0} & 
  		& \textbf{2}& \textbf{0} & 
  		& \textbf{0}& \textbf{0} & 
  		& \textbf{3}& \textbf{1} & \cite{d34} 
  		& \textbf{8}& \textbf{0} & 
  		\\
  
  & Internet
  		& \textbf{0}& \textbf{0} & 
  		& \textbf{0}& \textbf{0} & 
  		& \textbf{0}& \textbf{0} & 
  		& \textbf{7}& \textbf{0} & 
  		& \textbf{20}& \textbf{5} & \cite{a3}\cite{a4}\cite{a20} 
  		\\
  
  & Physical medium
  		& \textbf{0}& \textbf{0} & 
  		& \textbf{3}& \textbf{2} & \cite{a9}\cite{d32}\cite{a23}\cite{a24} 
  		& \textbf{0}& \textbf{0} & 
  		& \textbf{1}& \textbf{1} & \cite{a16}
  		& \textbf{4}& \textbf{4} & \cite{a1}\cite{a20}\cite{a21} 
  		\\
  
  & Independent
  		& \textbf{0}& \textbf{0} & 
  		& \textbf{0}& \textbf{2} & 
  		& \textbf{0}& \textbf{7} & \cite{a2}\cite{a10}\cite{a12}\cite{d8}\cite{d27}\cite{a18}\cite{a22} 
  		& \textbf{0}& \textbf{0} & 
  		& \textbf{3}& \textbf{0} & 
  		\\ \hline

  
  \multirow{4}{*}{\shortstack[l]{\textbf{Direct} \\ \textbf{Consequence}}} & DoS
  		& \textbf{0}& \textbf{0} & 
  		& \textbf{2}& \textbf{0} & 
  		& \textbf{0}& \textbf{0} & 
  		& \textbf{0}& \textbf{0} & 
  		& \textbf{2}& \textbf{1} & \cite{a21} 
  		\\ 
  
  & Privacy leakage
  		& \textbf{0}& \textbf{0} & 
  		& \textbf{0}& \textbf{0} & 
  		& \textbf{0}& \textbf{5} & \cite{a10}\cite{a12}\cite{d27}\cite{a18}\cite{a22} 
  		& \textbf{1}& \textbf{0} & 
  		& \textbf{15}& \textbf{1} & \cite{a11} 
  		\\ 
  
  & Function control
  		& \textbf{2}& \textbf{2} & \cite{a6}\cite{d28} 
  		& \textbf{2}& \textbf{5} & \cite{d29}\cite{a9}\cite{d32}\cite{a23}\cite{a24} 
  		& \textbf{0}& \textbf{3} & \cite{a2}\cite{d33}\cite{d8} 
  		& \textbf{9}& \textbf{4} & \cite{d29}\cite{a16}\cite{a17}\cite{d34} 
  		& \textbf{15}& \textbf{4} & \cite{a1}\cite{a3}\cite{a17} 
  		\\ 
  
  & Device control
  		& \textbf{1}& \textbf{0} & 
  		& \textbf{1}& \textbf{0} & 
  		& \textbf{0}& \textbf{0} & 
  		& \textbf{7}& \textbf{0} & 
  		& \textbf{19}& \textbf{6} & \cite{a1}\cite{a4}\cite{a20} 
  		\\ \hline

  
  \multirow{3}{*}{\shortstack[l]{\textbf{Technique} \\ \textbf{Scalability}}} & One
  		& \textbf{0}& \textbf{0} & 
  		& \textbf{1}& \textbf{2} & \cite{d29}\cite{d34} 
  		& \textbf{0}& \textbf{0} & 
  		& \textbf{3}& \textbf{2} & \cite{a16}\cite{a17} 
  		& \textbf{8}& \textbf{6} & \cite{a1}\cite{a4}\cite{a17}\cite{a20}\cite{a21} 
  		\\ 
  
  & Some
  		& \textbf{3}& \textbf{2} & \cite{a6}\cite{d28} 
  		& \textbf{4}& \textbf{3} & \cite{d32}\cite{a23}\cite{a24} 
  		& \textbf{0}& \textbf{8} & \cite{a2}\cite{d33}\cite{a10}\cite{a12}\cite{d8}\cite{d27}\cite{a18}\cite{a22} 
  		& \textbf{14}& \textbf{2} & \cite{d29}\cite{d34} 
  		& \textbf{33}& \textbf{6} & \cite{a3}\cite{a4}\cite{a11}\cite{a20} 
  		\\ 
  
  & Any
  		& \textbf{0}& \textbf{0} & 
  		& \textbf{0}& \textbf{0} & 
  		& \textbf{0}& \textbf{0} & 
  		& \textbf{0}& \textbf{0} & 
  		& \textbf{10}& \textbf{0} & 
  		\\ \hline

  
   \multirow{2}{*}{\textbf{Stealthiness}} & Unnoticeable
  		& \textbf{3}& \textbf{2} & \cite{a6}\cite{d28} 
  		& \textbf{3}& \textbf{0} & 
  		& \textbf{0}& \textbf{8} & \cite{a2}\cite{d33}\cite{a10}\cite{a12}\cite{d8}\cite{d27}\cite{a18}\cite{a22} 
  		& \textbf{14}& \textbf{2} & \cite{a17}\cite{d34} 
  		& \textbf{46}& \textbf{11} & \cite{a1}\cite{a3}\cite{a4}\cite{a11}\cite{a17}\cite{a20} 
  		\\ 
  
  & Environment change
  		& \textbf{0}& \textbf{0} & 
  		& \textbf{2}& \textbf{5} & \cite{d29}\cite{a9}\cite{d32}\cite{a23}\cite{a24} 
  		& \textbf{0}& \textbf{0} & 
  		& \textbf{3}& \textbf{2} & \cite{a16}\cite{d29} 
  		& \textbf{5}& \textbf{1} & \cite{a21} 
  		\\ \hline

  
  \multirow{3}{*}{\shortstack[l]{\textbf{Responsible} \\ \textbf{Disclosure}}} & Yes
  		& \textbf{3}& \textbf{0} & 
  		& \textbf{2}& \textbf{1} & \cite{d29}\cite{a23} 
  		& \textbf{0}& \textbf{1} & \cite{a2} 
  		& \textbf{9}& \textbf{2} & \cite{d29}\cite{a17} 
  		& \textbf{5}& \textbf{6} & \cite{a4}\cite{a17}\cite{a20} 
  		\\
  
  & No
  		& \textbf{0}& \textbf{0} & 
  		& \textbf{0}& \textbf{0} & 
  		& \textbf{0}& \textbf{0} & 
  		& \textbf{0}& \textbf{0} & 
  		& \textbf{4}& \textbf{0} & 
  		\\ 
  
  & Not specified
  		& \textbf{0}& \textbf{2} & \cite{a6}\cite{d28} 
  		& \textbf{3}& \textbf{3} & \cite{a9}\cite{d32}\cite{a24} 
  		& \textbf{0}& \textbf{7} & \cite{d33}\cite{a10}\cite{a12}\cite{d8}\cite{d27}\cite{a18}\cite{a22} 
  		& \textbf{8}& \textbf{2} & \cite{a16}\cite{d34} 
  		& \textbf{26}& \textbf{6} & \cite{a1}\cite{a3}\cite{a11}\cite{a21} 
  		\\ \hline

  
  \multirow{3}{*}{\shortstack[l]{\textbf{Vulnerability} \\ \textbf{Fix}}} & Yes
  		& \textbf{2}& \textbf{0} & 
  		& \textbf{2}& \textbf{1} & \cite{d29} 
  		& \textbf{0}& \textbf{1} & \cite{a2} 
  		& \textbf{7}& \textbf{1} & \cite{d29} 
  		& \textbf{23}& \textbf{2} & \cite{a20} 
  		\\ 
  
  & No
  		& \textbf{0}& \textbf{0} & 
  		& \textbf{0}& \textbf{0} & 
  		& \textbf{0}& \textbf{0} & 
  		& \textbf{1}& \textbf{0} & 
  		& \textbf{6}& \textbf{0} & 
  		\\
  
  & Not specified
  		& \textbf{1}& \textbf{2} & \cite{a6}\cite{d28} 
  		& \textbf{3}& \textbf{4} & \cite{a9}\cite{d32}\cite{a23}\cite{a24} 
  		& \textbf{0}& \textbf{7} & \cite{d33}\cite{a10}\cite{a12}\cite{d8}\cite{d27}\cite{a18}\cite{a22} 
  		& \textbf{8}& \textbf{3} & \cite{a16}\cite{a17}\cite{d34} 
  		& \textbf{20}& \textbf{10} & \cite{a1}\cite{a3}\cite{a4}\cite{a11}\cite{a17}\cite{a21} 
  		\\ \specialrule{.12em}{.05em}{.05em}

  \end{tabular}
\end{center}

\end{table*}

Our analysis on web reports also reveals the most reported consequences of IoT attacks. As shown in Figure~\ref{fig:year_direct_overall}, the increase in attacks can be mostly attributed to attacks which result in device control, function control and---at a lesser but also significant degree---privacy leakage. Device control includes exploits that enroll IoT devices into botnets. Given the recent massive DDoS attacks~\cite{mirai_million}, a lot of web reports have been focusing on that. Function control usually stems from a lack of authentication on devices which can be manipulated by an adversary. This is a low-effort attack since it does not require the adversary to get root control on the device; it only needs knowledge of the device APIs. Looking closer at those issues, we found that most of the affected entities are the IoT devices (A/S) themselves. We note that in those cases, again full device control and  function control are the prevalent consequences (see Figure~\ref{fig:role_direct_overall}). Increasingly, most attacks target home automation devices (see Figure~\ref{fig:year_deviceType_overall}). Wearable security was a popular topic in the media and industry in 2015 with significantly fewer reports in 2016. Not surprising, the most affected communication channels are WiFi and Internet. The former is the most popular way of communication between home automation A/S and a hub or router (D/R), while the Internet is what allows those devices to be remotely controlled (see Figure~\ref{fig:bar_channel_overall}). What this reveals is that \textit{there is an increasing interest in the security and privacy of home automation devices using WiFi and Internet.}

Next, we perform an in-depth assessment of all unique attack incidents reported by both industry and academia.



\subsection{Attack Incidents Analysis} \label{sec:attackdiscussion}

We organize our assessment on IoT attacks around the main problem areas we identify (see Section~\ref{sec:systemization} for definitions). This assessment is carried out based on the properties we defined earlier in this Section. Table~\ref{tab:attackMetrics} summarizes our analysis. Next, we discuss important observations stemming from this analysis.

\vspace{5pt}\noindent\textbf{LAN Mistrust}. This problem arises in attacks using WiFi as their communication channel and with the main target being IoT devices, hubs and routers. For example, in \cite{d28,trusthan1}, the adversary takes advantage of their position in the same network as the target IoT device. Due to the reliance on WiFi authentication, the target IoT devices do not take any extra measures to authenticate requests from devices on the same network, leaving them vulnerable to a local adversary. We further observe that this kind of attacks commonly achieve function control~\cite{d28} and are mostly stealthy~\cite{d28,trusthan1,trusthan2}. This happens because the adversary is implicitly trusted to use the control interface of the device resulting in expected operations but maybe in opportunistic moments. The threat model is as expected the LAN, since this problem manifests in connected smart home devices~\cite{d28,trusthan1,trusthan3}.
\begin{mybox}[boxsep=0pt,
                  boxrule=1pt,
                  left=4pt,
                  right=4pt,
                  top=4pt,
                  bottom=4pt,
                  ]
~Some IoT vendors trust devices on the same local network, leaving them vulnerable to local adversaries.
\end{mybox}
\vspace{-5pt}


\vspace{5pt}\noindent\textbf{Environment Mistrust}. This problem stems from proximity attackers which achieve function control~\cite{envi1, a9, a23, a24} and availability attacks~\cite{envi2} on their targets. In fact, 40\% (2/5) of the observed DoS attacks on devices are possible because of an implicit trust of the environment. In both cases, the attacks used close-range jamming signals to render their targets non-operational~\cite{envi2}. Such DoS attacks can be---according to our stealthiness definition---unnoticeable, since they do not affect a change to the environment. However, the user might eventually notice the offense since the device will be unresponsive. All other cases\ignore{~\cite{envi1,d29}} do affect the environment: for example, in \cite{envi1} an adversary can use a speaker which produces synthetic voice commands to control a smart TV, while \cite{d29} describes how a Bluetooth smart lock on the main entrance can unlock when the device owner enters home through the garage door. It is also evident from our analysis that at least 50\% of the attacks targeting user interaction point (UIP) can be linked to erroneous environment mistrust issue. This is partially due to attacks on emerging usable interfaces of IoT devices, such as those using voice and gesture signals. For example, users can now utilize voice assistants to control IoT devices (e.g. Google Now on Android phones, Cortana on Windows, Siri on iPhones and MACs and even Alexa on the Amazon Echo device). Despite advances in speech recognition, it seems that we lack good solutions to authenticate such acoustic signals, leaving the systems vulnerable to record and replay attacks. In fact, \cite{a9} demonstrates how acoustic signals emanating from a speaker, can be incomprehensible from a human but correctly interpreted and executed by such voice assistants.
\begin{mybox}[boxsep=0pt,
                  boxrule=1pt,
                  left=4pt,
                  right=4pt,
                  top=4pt,
                  bottom=4pt,
                  ]
~Home automation devices tend to suffer from environment mistrust problems stemming from emerging unauthenticated control signals, such as voice.
\end{mybox}
\vspace{-5pt}

\vspace{5pt}\noindent\textbf{App Over-privilege}. Interestingly, none of our collected web reports discuss attacks that fall within this problem area. Conversely, a large body of academic works exist. We believe this happens partially because these problems stem from more sophisticated adversaries which utilize side-channel techniques or perform inference attacks to exploit the failure of IoT systems and application platforms to enforce least privilege. Such attacks typically assume an adversary that can place a malicious app on either a user interaction point (UIP)~\cite{a10,d8,d27} or cloud server of an IoT hub~\cite{a2}. Such adversaries do not rely on a particular communication protocol and are typically stealthy. \cite{d8} focuses on the coarse granularity of the Android permission system which allows any app with a permission to surreptitiously access Bluetooth, NFC, Audio and Internet devices. Some malicious apps achieve manipulation of IoT devices' interfaces to instruct them to perform unauthorized operations. For example, in \cite{a2}, a malicious battery monitor SmartApp could subscribe door lock PIN change event by leveraging over-privilege problem on SmartThings platform. However, in most of the cases, the adversaries are passive, aiming to compromise data confidentiality. For example, a malicious app could use accelerometer, gyroscope and magnetometer on a smartwatch to infer ATM PIN key entries with more than 90\% accuracy after three trials~\cite{a18}.
\begin{mybox}[boxsep=0pt,
                  boxrule=1pt,
                  left=4pt,
                  right=4pt,
                  top=4pt,
                  bottom=4pt,
                  ]
~App over-privilege problem on smartphones is well studied in academia, however, it does not attract much attention from industry. It manifests in multi-application platforms (e.g. smartphones, hub) and is typically exploited by more sophisticated adversaries who perform side-channel and inference attacks.
\end{mybox}
\vspace{-5pt}

\vspace{5pt}\noindent\textbf{No/Weak Authentication}. End IoT devices typically suffer from authentication problems. For example, \cite{auth2} discusses how the Insteon hub does not authenticate by default remote requests, which allows an adversary to control any device connected to the hub. Almost 40\% of the cases involve a proximity adversary using Bluetooth to attack wearables~\cite{a17, d34, auth3, auth4} and other personal devices~\cite{auth5}. This is usually due to weak authentication mechanisms at the Bluetooth protocol level---devices either do not require a PIN~\cite{auth3} or allow a weak PIN that can be brute-forced~\cite{auth4}. To make things worse, vendors fail to implement authentication on the application layer~\cite{a17}. On the other hand, a remote adversary uses the Internet to attack problematic home automation devices~\cite{auth6}. These attacks oftentimes result in full control of the vulnerable IoT device. For example, researchers found that Ubi runs an ADB (Android Debug Bridge) service and a VNC service (providing access to the Android UI) with no password~\cite{auth1}. By leveraging this vulnerability, an attacker could gain root access on this device. Vehicles and their aftermarket IoT accessories also perform weak or no authentication. In \cite{a16} the adversary is assumed to tap into the controller area network (CAN bus) to control some of the car functions without authentication while in \cite{trusthan4} the adversary places an OBD dongle to control some of the target car's features. However, in both cases, the adversary is assumed to have physical access to the vehicle. A major problem here is the lack of isolation between electronic components in the vehicular network which share the same bus.
\begin{mybox}[boxsep=0pt,
                  boxrule=1pt,
                  left=4pt,
                  right=4pt,
                  top=4pt,
                  bottom=4pt,
                  ]
~No/weak authentication is a serious issue on consumer IoT devices. Bluetooth-enabled devices have been widely reported as vulnerable by industry. An area of interest attracting researchers from both industry and academia are authentication issues in emerging in-vehicle networks. This can be attributed to the severe consequences of potential attacks.
\end{mybox}
\vspace{-5pt}

\vspace{5pt}\noindent\textbf{Implementation Flaw}. Most of the attacks on IoT devices are due to implementation flaws. Adversaries on the same LAN can exploit open ports~\cite{impl1} to gain control of devices. Adversaries can also perform their attacks remotely exploiting either XSS vulnerabilities~\cite{a3}, SSL flaws~\cite{impl12} or \ignore{open ports~\cite{impl2}} hardcoded credentials~\cite{impl2}. For example, some vendors provide tutorials to users, guiding them to set up port forwarding to enable remote control~\cite{impl3}. However, if the devices are vulnerable, this allows remote attackers to exploit this attack surface to break into the device and in extend the local network. Unfortunately, a lot of IoT vendors follow this practice, which is problematic as we discuss later in~\ref{sec:discussion}. The LAN adversary model seems to be interesting only to the industry and the media while the remote model is studied by both. A reason for this could be the fact that remote adversary entails higher risk since it requires less effort and assumptions about the adversary's capabilities. In addition, implementation flaw problems seem to be the main focus for the industry and the media---more than 50\% of the reported attacks are possible due to implementation flaws. Such flaws manifest in home automation devices, such as light bulbs~\cite{impl4} and thermostats~\cite{impl5}, and are in a large extend attributed to hardcoded credentials in the devices' firmware~\cite{a4, impl6}, lack of validation of firmware updates~\cite{a20}, or unencrypted traffic that exposes credentials~\cite{auth1}. While, device-specific, these problems also result in severe consequences: In fact, most of the attacks which result in full device control (25/34) are due to implementation flaws. Moreover, such flaws enable scalable attacks: we found 9/10 attack cases that can be potentially applied to ``any device'', to be possible due to implementation flaws. For example, \cite{impl1} discovers that the website of a baby monitor device has a vulnerability which allows an adversary to login and view camera details of any other user. Interestingly, for a lot of reported attack cases, we do not find any vendor response. Note also that even if the vendor did respond with a software update, the update is not necessarily adopted by the users. A prominent example is the Mirai botnet \cite{impl9} which is linked to the recent massive DDoS attack on an Internet domain name server. Mirai exploits vulnerabilities in Internet-connected smart home devices to conscript them in a botnet and perform the DDoS attack. While the Mirai code was made available months ago~\cite{impl8}, and despite some vendor reaction (some offered security patches a year ago~\cite{impl10}), a lot of users never applied the security updates to their devices, allowing the attack to succeed. To make things worse, sometimes a software update cannot solve the problem, and vendors need to recall the vulnerable devices~\cite{impl7}. This is a process which might cost the manufacturer a lot of money and stigmatize their brand name; it also relies on users to make the effort to return the device. Lastly, we observe that all the attacks that do not report to vendors are exploiting implementation flaws. This might happen because the vulnerability founders aim to profit from them. For example, a cyber-security research firm who discovered implementation flaws in medical devices, partnered with an investment firm to profit by damaging the device vendor's stock price~\cite{impl11}.

Moreover, many IoT gadgets can be controlled by smartphone apps that may introduce new attack surfaces. Motivated by this, we perform static code analysis on 323 IoT apps crawled from Google Play. Our results are alarming: nearly 90\% of the apps trigger at least one security/privacy policy violation\ignore{defined by us}. For example, 44\% of the apps do not use encryption to protect their data, and about half the apps have at least one SSL/TLS related issue or do not use SSL at all.
Overall, the \textit{status quo} for IoT security is far from being optimistic even on the standard Android platform. One may expect the situation to be worse on custom platforms/OSes for IoT. Due to space constraints, the details of the analysis are presented in Appendix~\ref{sec:app_analysis}.

\begin{mybox}[boxsep=0pt,
                  boxrule=1pt,
                  left=4pt,
                  right=4pt,
                  top=4pt,
                  bottom=4pt,
                  ]
~One of the main sources of problems in IoT devices is implementation flaws. However, even when problems are reported or publicized, and vendors do provide a fix, this is not always applied by users. Hence, a lot of devices suffer from known issues, enabling attacks which can result in serious consequences such as full \textit{device control}.
\end{mybox}



\section{Security Solutions for IoT Systems}\label{sec:defense}
In scrutinizing the defenses for IoT, we identify various approaches. (a) Authentication (\textit{Auth}): these commonly employ app-level or end-to-end authentication. (b) \textit{Access control}: these works propose---and sometimes implement---access control schemes to improve management and security of IoT devices. Typically these approaches are independent of the IoT device software and hardware architecture; instead, they depend on a centralized entity (e.g. delegate/relay) to enforce access control policies. (c) Embedded Device Security (\textit{EDS}): these works mainly focus on buttressing the security of the IoT devices themselves by proposing new system architectures. Commonly these require specialized hardware since they borrow concepts from the trusted platform computing (TPM) domain to enable device attestation from a verifier. 
(d) Device Attestation (\textit{DA}): they are similar to EDS approaches but they do not necessarily rely on secure hardware. Instead, the verifier might utilize either the root of trust guaranteed provided by EDS or environment signals to attest to the integrity of messages rather than the integrity of the device hardware/software. (e) Intrusion Detection Systems (\textit{IDS}) : these approaches are more passive in nature and mainly focus on detection (traffic monitoring) rather than prevention. Commonly, they overhaul the network traffic/activity to detect a compromise. (f) Information flow (\textit{IF}): these approaches control the information flow based on different sensitivity levels. (g) \textit{Other}: approaches which are typically outside the computer and network realm (e.g. manufacturing/mechanical).

\ignore{
\begin{table}[]
\centering
\caption{Defense Approaches}
\label{tab:defenceTypes}
\begin{tabular}{l | l}
Approach                                                                   & Paper List \\ \hline
Device Attestation                                                         &            \\
Access Control                                                             &            \\
\begin{tabular}[c]{@{}l@{}}Traffic Monitoring \& \\ Firewalls\end{tabular} &            \\
Information Flow                                                           &           
\end{tabular}
\end{table}
}

In this section, we take a closer look into the IoT defense works, propose assessment properties and perform a security analysis focused on the problem areas they tackle.

\ignore{\subsection{Landscape}}




%
\begin{table*}[!htbp]
\caption{Assessment of Defenses for the Internet of Things}
\label{tab:defenceMetrics}
\scriptsize

\begin{center}
\begin{threeparttable}
  \def\arraystretch{1.3}
  \setlength\tabcolsep{0.9pt} 
  \begin{tabular}{ l r | M{0.41cm} M{0.39cm} M{0.39cm} | M{0.39cm} M{0.39cm} M{0.39cm} M{0.39cm} M{0.39cm} M{0.39cm} M{0.39cm} M{0.39cm} M{0.39cm} M{0.39cm} M{0.39cm} M{0.39cm} M{0.39cm} M{0.39cm} | M{0.39cm} M{0.39cm} | M{0.39cm} M{0.39cm} M{0.39cm} M{0.39cm} M{0.39cm} | M{0.39cm} M{0.39cm} | M{0.39cm} M{0.39cm} M{0.39cm} | M{0.55cm}}
  
  \toprule[0.12em]
  & & \multicolumn{3}{c|}{\textbf{Auth}} & \multicolumn{14}{c|}{\textbf{Access Control}} & \multicolumn{2}{c|}{\textbf{EDS}} & \multicolumn{5}{c|}{\textbf{DA}} & \multicolumn{2}{c|}{\textbf{IDS}} & \multicolumn{3}{c|}{\textbf{IF}} & \textbf{Other} \\  
  
  & & \cite{d34} & \cite{d22} & \cite{d35}
  	& \cite{d28} & \cite{d29} & \cite{d33} & \cite{d3} & \cite{d7}
  	& \cite{d8} & \cite{d9} & \cite{d15} & \cite{d17} & \cite{d19}
  	& \cite{d24} & \cite{d25} & \cite{d26} & \cite{d20} 
  	& \cite{d1} & \cite{d12} 
  	& \cite{d11} & \cite{d13} & \cite{d14} & \cite{d36} & \cite{d37}
  	& \cite{d27} & \cite{d6} 
  	& \cite{d5} & \cite{d38} & \cite{d39}
  	& \cite{d32}\\ \hline
  
  \multirow{5}{*}{\textbf{Problem Area}} & LAN mistrust  
  	& \epie & \epie & \epie
  	& \epie & \epie & \epie & \fpie & \epie 
  	& \epie & \epie & \epie & \epie & \epie 
  	& \epie & \epie & \epie & \fpie 
  	& \epie & \epie 
  	& \epie & \epie & \epie & \epie & \epie
  	& \epie & \epie 
  	& \epie & \epie & \epie
  	& \epie \\
  
  & Environment mistrust 
  	& \epie & \epie & \fpie
  	& \epie & \fpie & \epie & \epie & \epie 
  	& \epie & \epie & \epie & \epie & \epie 
  	& \epie & \epie & \epie & \epie  
  	& \epie & \epie 
  	& \epie & \hpie & \fpie & \epie & \epie
  	& \epie & \epie 
  	& \epie & \epie & \hpie 
  	& \fpie \\
  
  & App over-privilege 
  	& \epie & \epie & \epie
  	& \epie & \epie & \fpie & \epie & \epie 
  	& \fpie & \fpie & \epie & \epie & \epie 
  	& \epie & \epie & \epie & \epie 
  	& \epie & \epie 
  	& \epie & \epie & \epie & \epie & \epie
  	& \fpie & \epie 
  	& \fpie & \fpie & \fpie 
  	& \epie \\
  
  & No/Weak authentication 
  	& \fpie & \fpie & \epie
  	& \fpie & \epie & \epie & \fpie & \hpie 
  	& \epie & \fpie & \fpie & \fpie & \fpie 
  	& \fpie & \fpie & \fpie & \epie 
  	& \epie & \epie 
  	& \epie & \epie & \epie & \epie & \epie
  	& \epie & \hpie 
  	& \epie & \epie & \epie 
  	& \epie \\
  
  & Implementation flaw 
  	& \fpie & \epie & \epie
  	& \hpie & \epie & \epie & \hpie & \fpie 
  	& \epie & \epie & \hpie & \hpie & \hpie 
  	& \hpie & \hpie & \hpie & \epie 
  	& \hpie & \hpie 
  	& \hpie & \hpie & \epie & \hpie & \hpie
  	& \epie & \hpie 
  	& \epie & \epie & \epie 
  	& \epie \\ 
  
  \hline  
  
  \multirow{5}{*}{\textbf{Target Entity}} & IoT device (A/S)  
  	& \fpie & \epie & \epie
  	& \fpie & \fpie & \epie & \fpie & \fpie 
  	& \epie & \epie & \fpie & \fpie & \epie 
  	& \fpie & \fpie & \fpie & \fpie 
  	& \fpie & \fpie 
  	& \fpie & \fpie & \fpie & \fpie & \fpie
  	& \epie & \fpie 
  	& \epie & \epie & \epie 
  	& \fpie \\
  
  & Hub and router (D/R)
  	& \epie & \epie & \epie
  	& \epie & \epie & \epie & \fpie & \epie 
  	& \epie & \epie & \epie & \epie & \epie 
  	& \epie & \epie & \epie & \epie 
  	& \fpie & \fpie 
  	& \fpie & \epie & \epie & \fpie & \fpie
  	& \epie & \epie 
  	& \fpie & \epie & \epie 
  	& \epie \\
  
  & Cloud server (D/R)
  	& \epie & \epie & \epie
  	& \epie & \epie & \epie & \epie & \epie 
  	& \epie & \epie & \epie & \epie & \epie 
  	& \epie & \epie & \epie & \epie  
  	& \epie & \epie 
  	& \epie & \epie & \epie & \epie & \epie
  	& \epie & \epie 
  	& \fpie & \epie & \epie 
  	& \epie \\
  
  & Smartphone (UIP)
  	& \epie & \fpie & \fpie
  	& \epie & \epie & \fpie & \fpie & \epie 
  	& \fpie & \fpie & \epie & \epie & \fpie 
  	& \epie & \epie & \epie & \epie  
  	& \epie & \epie 
  	& \epie & \epie & \epie & \epie & \epie
  	& \fpie & \epie 
  	& \epie & \epie & \fpie 
  	& \epie \\
  
  & IoT device (UIP) 
  	& \epie & \epie & \epie
  	& \epie & \epie & \epie & \fpie & \epie 
  	& \epie & \epie & \epie & \epie & \epie 
  	& \epie & \epie & \epie & \epie 
  	& \fpie & \fpie 
  	& \fpie & \epie & \epie & \fpie & \fpie
  	& \epie & \epie 
  	& \epie & \fpie & \epie 
  	& \epie \\
  
  \hline

  \multirow{5}{*}{\textbf{Threat Model}} & Malicious app  
  	& \epie & \epie & \epie
  	& \epie & \epie & \fpie & \epie & \epie 
  	& \fpie & \fpie & \epie & \epie & \epie 
  	& \epie & \epie & \epie & \epie 
  	& \fpie & \fpie 
  	& \fpie & \epie & \epie & \fpie & \fpie
  	& \fpie & \epie 
  	& \fpie & \fpie & \fpie 
  	& \epie \\
  
  & LAN 
  	& \epie & \epie & \epie
  	& \epie & \epie & \epie & \fpie & \epie 
  	& \epie & \fpie & \epie & \epie & \epie 
  	& \epie & \epie & \epie & \fpie  
  	& \fpie & \fpie 
  	& \fpie & \fpie & \epie & \fpie & \fpie
  	& \epie & \fpie 
  	& \epie & \epie & \epie 
  	& \epie \\
  
  & Proximity 
  	& \fpie & \fpie & \fpie
  	& \epie & \fpie & \epie & \fpie & \fpie 
  	& \epie & \epie & \epie & \epie & \epie 
  	& \epie & \epie & \epie & \epie  
  	& \fpie & \fpie 
  	& \epie & \epie & \epie & \epie & \epie
  	& \epie & \epie 
  	& \epie & \epie & \fpie 
  	& \fpie \\
  
  & Remote 
  	& \epie & \epie & \epie
  	& \fpie & \epie & \epie & \fpie & \epie 
  	& \epie & \epie & \fpie & \fpie & \fpie 
  	& \fpie & \fpie & \fpie & \epie  
  	& \fpie & \fpie 
  	& \fpie & \fpie & \epie & \fpie & \fpie
  	& \epie & \epie 
  	& \epie & \epie & \epie 
  	& \epie \\
  
  & Physical 
  	& \epie & \epie & \epie
  	& \epie & \fpie & \epie & \epie & \epie 
  	& \epie & \epie & \epie & \epie & \epie 
  	& \epie & \epie & \epie & \epie 
  	& \fpie & \fpie 
  	& \epie & \fpie & \fpie & \fpie & \epie
  	& \epie & \epie 
  	& \epie & \epie & \epie 
  	& \epie \\
  
  \hline
  
  \multirow{6}{*}{\shortstack[l]{\textbf{Communication} \\ \textbf{Channel}}} & WiFi  
  	& \epie & \epie & \epie
  	& \fpie & \epie & \epie & \fpie & \epie 
  	& \epie & \fpie & \fpie & \fpie & \fpie 
  	& \fpie & \fpie & \fpie & X 
  	& \epie & \epie 
  	& \epie & \epie & \epie & \epie & \epie
  	& \epie & \epie 
  	& \epie & \epie & \epie 
  	& \epie \\
  
  & Internet 
  	& \epie & \epie & \epie
  	& \fpie & \epie & \epie & \fpie & \epie 
  	& \epie & \fpie & \fpie & \fpie & \fpie 
  	& \fpie & \fpie & \fpie & X 
  	& \epie & \epie 
  	& \epie & \fpie & \epie & \epie & \epie
  	& \epie & \epie 
  	& \epie & \epie & \epie 
  	& \epie \\
  
  & Bluetooth 
  	& \fpie & \fpie & \epie
  	& \epie & \fpie & \fpie & \fpie & \fpie 
  	& \fpie & \epie & \epie & \fpie & \epie 
  	& \epie & \epie & \epie & X 
  	& \epie & \epie 
  	& \epie & \epie & \epie & \epie & \epie
  	& \epie & \epie 
  	& \epie & \epie & \epie 
  	& \epie \\
  
  & Physical 
  	& \epie & \epie & \fpie
  	& \epie & \epie & \epie & \epie & \epie 
  	& \epie & \epie & \epie & \epie & \epie 
  	& \epie & \epie & \epie & X 
  	& \epie & \epie 
  	& \epie & \epie & \fpie & \epie & \epie
  	& \epie & \epie 
  	& \epie & \epie & \fpie 
  	& \fpie \\
  
  & Independent 
  	& \epie & \epie & \epie
  	& \epie & \epie & \epie & \epie & \epie 
  	& \epie & \epie & \epie & \epie & \epie 
  	& \epie & \epie & \epie & X 
  	& \fpie & \fpie 
  	& \fpie & \epie & \epie & \fpie & \fpie
  	& \fpie & \epie 
  	& \fpie & \fpie & \fpie 
  	& \epie \\
  	
  & Others 
  	& \epie & \epie & \epie
  	& \epie & \epie & \epie & \fpie & \epie 
  	& \fpie & \epie & \epie & \fpie & \epie 
  	& \epie & \epie & \epie & X 
  	& \epie & \epie 
  	& \epie & \epie & \epie & \epie & \epie
  	& \epie & \fpie 
  	& \epie & \epie & \epie 
  	& \epie \\
  
  \hline

  \multirow{4}{*}{\textbf{Modification}} & Hardware  
  	& \epie & \epie & \epie
  	& \epie & \fpie & \epie & \epie & \epie 
  	& \epie & \epie & \epie & \epie & \epie 
  	& \epie & \epie & \epie & X  
  	& \fpie & \fpie 
  	& \fpie & \epie & \epie & \fpie & \fpie
  	& \epie & \epie 
  	& \epie & \epie & \epie 
  	& \fpie \\
  
  & New hardware 
  	& \epie & \epie & \epie
  	& \fpie & \epie & \epie & \epie & \fpie 
  	& \epie & \epie & \fpie & \fpie & \fpie 
  	& \fpie & \fpie & \fpie & X 
  	& \epie & \epie 
  	& \epie & \fpie & \epie & \epie & \epie
  	& \epie & \fpie 
  	& \epie & \epie & \epie 
  	& \epie \\
  
  & OS update 
  	& \fpie & \fpie & \fpie
  	& \epie & \epie & \fpie & \fpie & \epie 
  	& \fpie & \fpie & \epie & \epie & \epie 
  	& \epie & \epie & \epie & X 
  	& \epie & \epie 
  	& \epie & \epie & \epie & \epie & \epie
  	& \epie & \epie 
  	& \fpie & \fpie & \fpie 
  	& \epie \\
  
  & Application 
  	& \epie & \epie & \epie
  	& \epie & \epie & \epie & \epie & \epie 
  	& \epie & \epie & \epie & \epie & \epie 
  	& \epie & \epie & \epie & X 
  	& \epie & \epie 
  	& \epie & \epie & \fpie & \epie & \epie
  	& \fpie & \epie 
  	& \epie & \epie & \epie 
  	& \epie \\
  
  \hline
  
  \multicolumn{2}{l|}{\textbf{Backward Compatibility}}  
  	& \hpie & \fpie & \fpie
  	& \fpie & \epie & \fpie & \hpie & \fpie 
  	& \fpie & \hpie & \fpie & \fpie & \hpie 
  	& \fpie & \fpie & \fpie & X  
  	& \epie & \epie 
  	& \epie & \fpie & \fpie & \epie & \epie
  	& \fpie & \fpie 
  	& \fpie & \fpie & \fpie 
  	& \epie \\
  
  \hline
  
  \multirow{3}{*}{\shortstack[l]{\textbf{User} \\ \textbf{Interaction}}} & No  
  	& \epie & \fpie & \epie
  	& \epie & \fpie & \fpie & \epie & \epie 
  	& \epie & \epie & \epie & \epie & \epie 
  	& \epie & \epie & \epie & \epie  
  	& \fpie & \fpie 
  	& \epie & \epie & \epie & \epie & \epie
  	& \epie & \epie 
  	& \epie & \epie & \epie 
  	& \fpie \\
  
  & Bootstrap 
  	& \fpie & \epie & \fpie
  	& \fpie & \epie & \epie & \fpie & \fpie 
  	& \fpie & \fpie & \fpie & \fpie & \fpie 
  	& \fpie & \fpie & \fpie & \fpie  
  	& \epie & \epie 
  	& \epie & \epie & \epie & \epie & \epie
  	& \epie & \epie 
  	& \fpie & \fpie & \epie 
  	& \epie \\
  
  & Prompt 
  	& \epie & \epie & \epie
  	& \epie & \epie & \epie & \epie & \epie 
  	& \epie & \epie & \epie & \epie & \epie 
  	& \epie & \epie & \epie & \epie  
  	& \epie & \epie 
  	& \fpie & \fpie & \fpie & \fpie & \fpie
  	& \fpie & \fpie 
  	& \epie & \epie & \fpie 
  	& \epie \\
  
  \toprule[0.12em]

  \end{tabular}

\begin{tablenotes}
\item[] \fpie\ = fully applies; \hpie\ = applies in some cases; \epie\ = does not apply; X\ = not specified.
\end{tablenotes}
\end{threeparttable}
\end{center}

\end{table*}

\subsection{Assessment Properties for Defenses}
To better understand the proposed defenses by researchers, we utilize the \textbf{problem area} (see Section~\ref{sec:taxonomy}), and the \textbf{target entity}, \textbf{threat model} and \textbf{communication channel} properties (see Section~\ref{sec:attackmetrics}). In addition to these, we further define properties which as particular to IoT defense approaches, namely \textbf{modification}; \textbf{backward compatibility} and; \textbf{user interaction}. Next, we define these properties.

\vspace {5pt}\noindent\textbf{Modification.} The modification property aims to capture the hardware and software requirements of the proposed solution. In terms of hardware, a solution might require either (a) changes to the \textit{hardware} of the existing system entities or (b) the addition of \textit{new hardware} into the environment. For example, a TPM approach for an IoT device will require a secure chip on the existing hardware, while a framework to control access to various IoT devices, might require the addition of a new hub which has support for various communication channels. In terms of software, a solution might require (c) an \textit{OS update} (e.g. router firmware changes) or (d) \textit{application} changes (e.g. app installation on a user's smartphone).

\vspace {5pt}\noindent\textbf{Backward Compatibility.} This property aims to describe whether a solution is backward compatible or not. We regard a solution as backward compatible if when the solution is applied, the target devices gain the protection benefits without the need of replacing the devices themselves. Some systems could be partially backward compatible if an update is needed but can be applied over the air.


\vspace {5pt}\noindent\textbf{User Interaction.} Some solutions require (a) \textit{no} user interaction; (b) interaction only to \textit{bootstrap} the system; or (c) \textit{prompt} users for making decisions. For example, a TPM solution for an IoT device is typically completely transparent to users. On the other hand, some access control schemes and traffic monitoring techniques require users to setup the access control policies once to bootstrap the system; others prompt users to make decisions during enforcement.




\subsection{Proposed Solutions Analysis}\label{sec:defensediscussion}
We categorize the defenses according to the problem areas they solve. We utilize the defined properties to perform an in-depth assessment of the approaches used and summarize our results in Table~\ref{tab:defenceMetrics}. Next, we discuss the main insights drawn from this assessment.

\vspace{5pt}\noindent\textbf{LAN Mistrust}. Two works can solve this problem for the systems they consider~\cite{d3,d20}. \cite{d20} focuses on a home environment where a visitor who is authenticated to the network attempts to compromise devices on the local network. The authors perform a user study to determine access control policies for LAN devices. However, there is no discussion on how these can be implemented and enforced and as such we could not assign values to most of the properties. \cite{d3} proposes an end-to-end authentication and capability-based access control model which can thwart attacks from LAN, proximity and remote adversaries. However, every device needs to update its firmware and be authenticated to other devices it wants to communicate with. If a user owns several devices, setting up all devices and assigning the correct capabilities could be a laborious endeavor. 
\begin{mybox}[boxsep=0pt,
                  boxrule=1pt,
                  left=4pt,
                  right=4pt,
                  top=4pt,
                  bottom=4pt,
                  ]
~End-to-end authentication and access control are both effective in solving \textit{LAN Mistrust} issues. However, the former relies on a universal implementation of the protection by all IoT vendors. Lastly, we note the absence of practical access control solutions that solve the \textit{LAN Mistrust} issue.
\end{mybox}
\vspace{-5pt}

\ignore{The takeaway from our analysis is that an access control approach might be effective in solving such issues. However, we lack practical solutions for either home networks or in-vehicle networks.}

\vspace{5pt}\noindent\textbf{Environment Mistrust}. Various approaches have been proposed to solve environment mistrust issues. Two out of five are device attestation solutions. \cite{d13} uses visual challenges (e.g. QR codes) to verify the freshness of camera footage. The goal is to detect adversaries who have tampered the camera to provide obsolete videos. The proposed solution only partially solves this problem: an attacker could bypass the system by physically moving or rotating the camera to keep it away from the target area, while at the same time keeping the visual challenge in sight. It further requires the addition of at least one new device (a verifier) which verifies the challenge. \cite{d14} does focus on physical adversaries who tamper WiFi IoT devices by moving or rotating them. The solution leverages the 802.11n WiFi frame preamble to detect tamper events. Unfortunately, their technique, while easy to deploy (just a software update) is not very effective: it is only proven to detect 53\% of the tamper events. Moreover, solutions following a device attestation approach commonly need to prompt the user on any detected event. The access control solution~\cite{d29} aims to protect Bluetooth-enabled smart locks from a physical attacker or an attacker in proximity. The proposed solution requires hardware changes to the smart locks and a new wearable device to be carried by the users. This renders it hard to deploy. \cite{d39} proposes an information flow approach to solve attacks using malicious voice commands. However, it is vulnerable to voice replay attacks. \cite{d35} proposes a liveness detection system for voice authentication on smartphones, working against replay\ignore{pre-recorded} voice attacks. It utilizes the time-difference-of-arrival changes to the two microphone of the smartphone when a human speaks. While effective, it does require the phone to be placed at a specific position. Lastly,~\cite{d32} tries to prevent privacy leakage from head mounted displays (e.g. Google Glass). The paper proposes adding a polarized lens in front of the display to prevent its reflection which can be potentially interpreted by a proximity adversary. We observe that most of the solutions in this space focus on proximity or physical adversaries since there is an implicit relationship between the capabilities of such adversaries and the problems they can induce to IoT systems. \ignore{Dependent on the particular application, approaches might vary, but typically device attestation solutions are more applicable to physical adversaries, they are backward compatible but tend to frequently prompt the user.}
\begin{mybox}[boxsep=0pt,
                  boxrule=1pt,
                  left=4pt,
                  right=4pt,
                  top=4pt,
                  bottom=4pt,
                  ]
~Environment mistrust problems are very specific to IoT devices which can be placed in diverse physical environments. Different approaches can be followed to solve the problem but are highly dependent on the application. Complete solutions need to consider physical adversaries to be effective.
\end{mybox}
\vspace{-5pt}

\vspace{5pt}\noindent\textbf{App Over-privilege}. This problem typically arises due to the coarse granularity of the protection mechanisms of entities that support multiple applications. These could be \textit{UIP}s (e.g. smartphones) or \textit{D/R}s (e.g. hubs), which support third-party apps. Three main approaches have been followed to mitigate this kind of problems. Dabinder~\cite{d33}, SEACAT~\cite{d8} and Xapp~\cite{d9} propose access control solutions. In particular \cite{d8} proposes modifications to the Android OS on smartphones, to support application-level access control to Bluetooth, NFC and audio devices in the vicinity. AppGuardian~\cite{d27} detects and prevents privacy being leaked from connected devices or other apps to malicious apps. It achieves that by blocking apps which continuously poll system resources in an attempt to gain inference through side-channel information. In contrast with the access control approaches, it requires no system modifications---the solution is deployed in the form of a third-party app. FlowFence~\cite{d5} on the other hand targets emerging IoT application frameworks. It proposes an information flow approach to prevent over-privilege third-party apps from misusing the end IoT devices of the framework. We observe that access control approaches require some user involvement to setup the policy rules but after that enforcement happens transparently. On a par with Dabinder and SEACAT, FlowFence also needs user aid in setting up the policies: the user needs to approve the developer issued flow rules. \cite{d38, d39} also propose information flow approaches that separate highly sensitive information from less sensitive information and apply access control to each flow. AppGuardian, as an IDS approach, might continuously prompt the user if an app is frequently accessing system resources and cannot be killed. We also note that all of the approaches in solving the app over-privilege problem can be effective and backward compatible. An exception is ~\cite{d9} which assumes that participating devices use a modified version of the OSGi framework.
\begin{mybox}[boxsep=0pt,
                  boxrule=1pt,
                  left=4pt,
                  right=4pt,
                  top=4pt,
                  bottom=4pt,
                  ]
~App over-privilege problems generally stem from design issues (e.g.failure to achieve the least privilege principle). Therefore, they typically require vendors to significantly modify the system in order to eradicate the problem.
\end{mybox}
\vspace{-5pt}

\vspace{5pt}\noindent\textbf{No/Weak Authentication}. A straightforward way to address these problems is to add end-to-end or application-level authentication. MASHaBLE~\cite{d22} and FitLock~\cite{d34} follow such approaches. They both target Bluetooth devices and as such, they consider a proximity threat model. They are also backward compatible but have different levels of user interaction. FitLock asks the user to setup the device in her account and enters a generated code for validation. In contrast, MASHaBLE enables cryptographic secret handshakes for mutual authentication among BLE supported devices (e.g. smartphones). Such approaches are indeed effective in solving the problem, however, they are application specific which delegates trust to app developers. Given the plethora of problems stemming from implementation flaws, relying solely on app developers for securing IoT could be catastrophic. 

Conversely, \cite{d28, d3, d7, d9, d15, d17, d19, d24, d25, d26} follow an access control approach to effectively tackle no/weak authentication problems. In this cases, trust is moved from IoT app developers to the D/Rs and UIPs, effectively reducing the trusted computing base of the ecosystem. An exception is ~\cite{d7} which only partially solves the problem. While it offers device-level authentication, it fails to guarantee app-level authentication. Most of these solutions focus on IoT devices that can be controlled through the Internet (remotely) and WiFi (locally) (9/10). Some (3/10) can solve authentication issues on Bluetooth devices: IACAC~\cite{d3} is a capability-based access control scheme implemented on WiFi but can be extended to the Bluetooth protocol. Thus to be deployed, the participating devices will need to update their protocol software stacks. \cite{d17, d26} extends the OSGi framework to built an access control mechanism supporting various communication protocols such as X10, Insteon, Zigbee, UPnP. The OSGi (Open Service Gateway Initiative) framework was used by various works~\cite{d9,d17,d26}. This is a modular service platform for the Java programming language which while proposed for gateways, it can be used in other devices such as smartphones and automobiles. Its extensible structure allows for over the air updates and integration of heterogeneous protocols. On the other hand, it is not widely deployed, and as such it would require firmware updates on all participating devices. 

\ignore{
Interestingly, \cite{d9,d19} focus on a LAN threat model but they do not solve the \textit{LAN Mistrust problem}---they are instead effective in tackling more general authentication problems on smartphones. \cite{d9} requires the device to employ adequate sandboxing and privilege separation, and to run a number of new modules. However, most small sensor devices do not have such capabilities. \cite{d19} considers smartphones exposed in an insecure public Wifi.
}

Lastly, \cite{d6} is an intrusion detection approach for a vehicular network. In particular, the paper considers a compromised ECU (electronic control unit) connected to the CAN (controller area network) bus of the car. It uses the intervals of periodic messages on the CAN bus as a signal to profile ECU units and detect anomalies. This can help detect issues in in-vehicle networks stemming from authentication problems but cannot prevent damage.
\begin{mybox}[boxsep=0pt,
                  boxrule=1pt,
                  left=4pt,
                  right=4pt,
                  top=4pt,
                  bottom=4pt,
                  ]
~In general, access control approaches are effective in tackling no/weak authentication problems and are independent of IoT vendors. However, they do require users to be involved in bootstrapping the system by helping with the setup of access control policies.
\end{mybox}
\vspace{-5pt}

\vspace{5pt}\noindent\textbf{Implementation Flaws}. Most of the access control solutions, apart from being effective in tackling no/weak authentication problems, they can also partially solve implementation flaws, for example, open port flaws. An access control solution such as~\cite{d26} can protect the device since it will block remote requests from unauthorized parties. In fact most of these solutions, focus on a remote adversary~\cite{d28,d15,d17,d24,d25,d26} which performs the attack across the Internet. This is incomplete since they seem to disregard, local threats such as malicious apps running on smartphones. For example, \cite{d25} assigns to an access point located one hop away from the target IoT device, the responsibility of securely tunneling the traffic to mitigate privacy leakage. In this case, if the device erroneously transfers sensitive information in plaintext (e.g. through HTTP), the solution will effectively tackle a remote adversary who passively eavesdrops on the traffic. However, a local threat might eavesdrop on the link between the device and the first hop compromising the confidentiality of the data in transit. Some other access control solutions can be more effective in tackling implementation flaws. For example, BLE-Guardian~\cite{d7} protects BLE devices by employing reactive jamming to hide BLE advertisements of these devices. This approach is effective for BLE: even if devices have implementation flaws, an external adversary cannot even discover the vulnerable device. On the other hand, on a par with \cite{d33}, a co-installed malicious app on the BLE device could potentially succeed.

Fitlock~\cite{d34} redesigns binding and uploading data procedures on a Bluetooth device (Fitbit), and provides an end-to-end authentication solution. This approach can solve implementation flaws on the IoT device when facing a proximity attacker. However, the firmware of the IoT device needs to be updated to adopt the protection.

Both \cite{d1, d12} employ an EDS approach. While they can protect against very powerful adversaries (see threat model), they can only partially solve implementation flaws. For example, even if an adversary exploits implementation flaws, these approaches would prevent system-level tampering, however, they can neither detect nor prevent attacks exploiting open ports to gain app-level function control.\ignore{The latter exploits application-level issues and does not need to alter the system or the apps in any way---it merely misuses the existing app APIs.} Moreover, EDS approaches can be applied to any device, whether this is an A/S, a D/R or a UIP. They are also mostly independent of communication channels since they focus on device security and not communication security. In addition, they act transparently to the user. However, they do require special hardware and as such are not always backward compatible. In general, EDS approaches can provide the means or tools to applications to attest to the integrity of the platform or themselves. They can prevent an exploit from escalating its privileges to system privileges, but the apps are responsible for their own security. 


Device attestation approaches also partially solve implementation flaws but do not necessarily require hardware modifications. For example, SEDA~\cite{d11} and SANA~\cite{d36} can effectively attest IoT device swarms against software attacks. Lastly, the IDS approaches usually facilitate detection of a compromise, but only after it has taken place.

In essence, although a lot of academic works can solve some implementation issues with IoT devices, this is not their main focus. This is partially because these issues are not very interesting from an academic perspective. These works opt in studying more fundamental design issues as discussed before. Nevertheless, implementation flaws are a non-trivial problem in IoT and we need practical solutions. An easy fix would be to update the vulnerable target entities. This assumes at least that the vulnerability is known to the vendor and that an over the air update is supported. The problem here is that these are not always valid assumptions and that such an approach relies on a diverse set of multiple IoT vendors who do not always treat security as a priority.
\begin{mybox}[boxsep=0pt,
                  boxrule=1pt,
                  left=4pt,
                  right=4pt,
                  top=4pt,
                  bottom=4pt,
                  ]
~Implementation flaws are a non-trivial problem in IoT and we lack practical, device independent and backward compatible solutions.
\end{mybox}


\section{Suggestions for IoT Security}\label{sec:discussion}



In the previous sections, we elaborate our study on the existing attacks on IoT systems (Section~\ref{sec:attack}) and the defenses against such threats (Section~\ref{sec:defense}). Based on the observations obtained from the study, here we present what we have learned and what needs to be done in this new area. 
 

\subsection{OBSERVATION: Emerging IoT Application Platforms Become New Attack Targets}

Our study highlights the emerging threats to the vulnerable delegate/relay targets (20/107). These targets are the IoT application platforms developed by IoT industry, such as Samsung's SmartThings~\cite{smartthings_dev}, Weave~\cite{weave_dev} and MiOS~\cite{mios_dev}. Although so far there is little evidence that indeed these platforms are exploited in the real world---mainly because these systems are relatively in their infancy---still the threats they are facing are real. Just as potentially harmful apps have already become the main threat to mobile platforms (Android and iOS), it is imaginable that in the near future, the IoT platforms can easily fall victim to similar attacks. In fact, recent studies~\cite{a2} show the possibility of the attacks from over-privileged applications on those platforms. However, based on what we have collected, apparently, these emerging threats have not yet been extensively investigated.
\begin{mybox}[boxsep=0pt,
                  boxrule=1pt,
                  left=4pt,
                  right=4pt,
                  top=4pt,
                  bottom=4pt,
                  ]
\textit{SUGGESTION:} More extensive and thorough studies are expected on the security of IoT application platforms, helping better understand their weaknesses and enhance their protection (isolation, access control) before real attacks happen, without undermining their utilities.
\end{mybox}


\subsection{OBSERVATION: No Protection for IoT User Interaction Points}

The most common way to control IoT devices is through the user's smartphone. However, in an attempt to streamline the process, more and more vendors provide other channels to help the user better manage her devices. A prominent example is Amazon Echo, a connected device which accepts an acoustic signal, translates it into commands (just like the smartphone services such as Google Now, Siri and Cortana) and further relays them to the corresponding device's cloud service. Another example is the gesture based control~\cite{ring,bird,myo} being developed by vendors.  The use of these techniques, however, also exposes IoT systems to new risks: if such physical communication channels (e.g., voice, gesture) are not properly mediated, an adversary could issue arbitrary commands to control connected devices. As an example, a study~\cite{a9} already demonstrates how a malicious device could construct and issue acoustic signals which while incomprehensible by humans, are accepted by voice assistants. These emerging challenges could render Environment Mistrust problems increasingly important for IoT. However, little has been done to understand how to protect such systems against such threats.

\begin{mybox}[boxsep=0pt,
                  boxrule=1pt,
                  left=4pt,
                  right=4pt,
                  top=4pt,
                  bottom=4pt,
                  ]
\textit{SUGGESTION:} Threats from the physical environment of IoT systems are understudied. The security risks of new control interfaces (e.g., voice, gesture) need to be better understood and effectively protected. 
\end{mybox}



\subsection{OBSERVATION: Lack of Security Solutions for Connected Cars}

According to Business Insider, the sale of connected cars is expected to grow rapidly, with estimated 45\% of the cars on the road to be connected by 2020~\cite{connect_car_report}. Google, Ford, Uber and Tesla are among the most prominent players experimenting with various levels of autonomous driving. In order to enable such capabilities, a large number of sensors and electronic components will be extensively deployed, which opens new attack surfaces with significant impacts. As an example, a prior study~\cite{a17,a20} reveals vulnerabilities in aftermarket dongles, which allow an adversary to assume control of vehicles.  For another example, ~\cite{a16} shows how an adversarial or compromised ECU unit can attack other components connected to the CAN bus of legacy vehicles.  More recently, even next generation vehicles (e.g. Tesla) are found to be vulnerable~\cite{tesla_hack}. 
Evidently, the existing in-vehicle computer and network infrastructure are not yet ready to guarantee the security of their components. Vulnerabilities in vehicles are serious and can even cost human lives. Our work shows that
more needs to be done to mitigate the threats to vehicular sensors, networks and architectures.
\begin{mybox}[boxsep=0pt,
                  boxrule=1pt,
                  left=4pt,
                  right=4pt,
                  top=4pt,
                  bottom=4pt,
                  ]
\textit{SUGGESTION:} In the next few years connected cars will be all around us. This highlights the importance of studying the security of electronic components in vehicular networks. A possible direction would be the design of new in-vehicle network architectures that isolate safety critical components from infotainment and other functions. Any solution here should guarantee the availability of security critical components, the integrity of their messages and authentication of remote commands.
\end{mybox}

\subsection{OBSERVATION: Pervasiveness of Implementation Flaws and No/Weak Authentication}

Our analysis reveals a significant number of attacks exploiting either implementation flaws or no/weak authentication issues in commercial IoT devices. In fact, 62/103 of all attacks are made possible because of the former and 19/103 because of the latter. To make things worse, such attacks can have severe impacts on the society: the Mirai botnet exploits such flaws in online devices to enroll them in large scale DDoS attacks. A version of Mirai recently conducted a massive DDoS attack in the US, rendering high profile online services (e.g. Twitter, Netflix, Amazon, PayPal) unavailable for hours. Our analysis on defense shows that there is no panacea for these kinds of problems. 

The manifestation and plethora of implementation flaws demonstrate that the whole IoT industry is yet to mature, with most products produced by start-ups, which oftentimes have little resource and incentive to build well thought-out security protection. In fact, we found that out of 357 home automation companies who provide statistical data, more than $3/4$ (167/217) are small businesses with less than 10 employees, 42/217 have between 11-50 and less than 5\% (7/217) have more than 50, which indicates the immaturity of this domain. Given the stringent competition, delivering a product/feature is often the top priority.  Unfortunately, security commonly pays the price due to the lack of or negligent code reviews, or in most of the cases lack of a security vision/plan in the development process.
\begin{mybox}[boxsep=0pt,
                  boxrule=1pt,
                  left=4pt,
                  right=4pt,
                  top=4pt,
                  bottom=4pt,
                  ]
\textit{SUGGESTION:} Industry experts, legislators and academics should come together to design a secure framework for consumer IoT devices. The existence of such a framework with security standards that IoT vendors have to adhere by would help reduce the trivially and massively exploitable vulnerabilities.
\end{mybox}




\subsection{OBSERVATION: Access Control Solutions Are Effective but Incomplete}

While a secure framework solution would go a long way in eradicating most of the implementation flaws, there will always be devices that lack protection. In fact, a lot of solutions proposed on academic papers seem to shift responsibility to the delegate/relay (either the router, cloud or local hub) which takes on the challenge of integrating a diverse set of devices with heterogeneous protocols in a common management and access control framework. These approaches reduce the trusted computing base since they are independent of the IoT vendor practices. Moreover, this independence allows them to effectively protect access on devices, irrespective of the presence of implementation flaws in them. Nevertheless, our study on such access control based defense reveals that the existing access control approaches commonly do not solve the \textit{LAN Mistrust} problem. This is because the vast majority of them are restricted to a remote adversary, taking no measure against internal (within LAN) threats. For example, an IoT device itself could be compromised or malicious by design~\cite{a4}. In addition, smartphones carry apps from multiple and not always verified sources. Thus a malicious app on an authorized phone might attack any device on the LAN. Our study on attacks already illustrates the existence of such LAN Mistrust attacks.
\begin{mybox}[boxsep=0pt,
                  boxrule=1pt,
                  left=4pt,
                  right=4pt,
                  top=4pt,
                  bottom=4pt,
                  ]
\textit{SUGGESTION:} Traditional LAN network solutions are too coarse-grained. There is a lack of fine-grained, IoT device-independent solutions. Access control defense employed within the LAN can solve the \textit{LAN Mistrust} and \textit{No/Weak Authentication} problems and mitigate most of the attacks exploiting \textit{implementation flaws}. Nevertheless, while academia has been studying access control schemes for years, we need new approaches tailored to the IoT reality: an access control scheme should take into account the existence of malicious apps on smartphones and/or integration hubs, and the possibility of malicious IoT devices within the LAN. At the same time, any solution should be backward-compatible: it is impractical to assume that all existing devices can be upgraded to seamlessly work with the new router/hub.
\end{mybox}



\section{Related Work}\label{sec:related}



\cite{journals/cn/SicariRGC15} and~\cite{journals/winet/JingVWLQ14} emphasized the layered security architecture analysis and available security solutions, respectively. Other works, like~\cite{medaglia2010overview, journals/computer/RomanNL11, journals/wpc/HeerMHKKW11}, appeared at the early stage of the mobile era before the explosion of IoT device ownership. Therefore, they inevitably miss current security challenges. Also, most of these works are short of real-world IoT attack cases to support their discussions. Instead, our collected attack incidents from multiple sources could offer the security community a more complete understanding of the challenges the IoT ecosystem is facing today. Additionally, the sound assessment properties we designed could systemize future reviews on IoT threats and solutions.\ignore{ rather than talk general security criterions one by one. ??}

IoT security involves many elements, and every single one could be a separate research topic. Granjal et al.~\cite{journals/comsur/GranjalMS15} focused on existing protocols and mechanisms to secure communications in IoT. Singh et al. \cite{journals/iotj/SinghPBKE16} offered security considerations for the IoT
architecture considering participating cloud services. \cite{report/Jason16} presents a policy report on IoT; Yan et al.~\cite{journals/jnca/YanZV14} delivered a survey on IoT trust management technologies, and Mukherjee~\cite{journals/pieee/Mukherjee15} is mainly concerned with physical-layer security, especially the resource-constrained secrecy coding and secret-key generation. Moreover, Denning et al.~\cite{journals/cacm/DenningKL13} reviewed potential computer security attacks against in-home technologies and presented a framework for the security risk evaluation. Weber~\cite{weber2010internet} studied the legal challenges posed by the secure implementation of IoT architecture. On the aspect of usable security, Ur et al.~\cite{ur2013current} found that smart home devices have their own isolated access control and fail to support user understanding for home access control. In contrast with these works and instead of focusing on a single attack surface in IoT, this SoK performs a comprehensive study on reported IoT attacks and defense to identify the major security and privacy problem areas in the space.




Other works focus on security issues of particular IoT devices or architectures. For instance, Rushanan et al.~\cite{conf/sp/RushananRKS14} surveyed publications related to security and privacy in implantable medical devices (IMDs) and health-related body area networks (BANs). Sadeghi et al.~\cite{conf/dac/SadeghiWW15} pointed out the security risks on Industrial IoT systems, and Wang et al.~\cite{journals/cn/WangL13} presented a survey of cyber security threats against the Smart Grid. 

Other related surveys on IoT security and dependability include wearable devices~\cite{journals/tmscs/AriasWHJ15}, IoT routing~\cite{journals/jnca/AirehrourGR16}, smart homes~\cite{journals/comsur/KomninosPP14}, embedded system design~\cite{journals/tecs/RaviRKH04}, etc. For reviews in the general domain of IoT, we refer the interested readers to the solid works of Atzori et al.~\cite{journals/cn/AtzoriIM10} and Perera et al.~\cite{journals/comsur/PereraZCG14}.


\section{Conclusion}\label{sec:conclusion}

In this paper, we report our study on the state of the art in consumer-facing IoT security and privacy. To provide a comprehensive picture of the environment and capture both current practices and new threats and defenses, we collected not only relevant academic papers but also thousands of industry reports in the form of white papers, news articles and blog posts. To serve this purpose, we built a literature retrieval and mining tool which gathers relevant online data, extracts semantic information and clusters similar articles together to form stories. We scrutinized the most reported attack incidents from the industry and the academic research and propose a number of properties which we use to systematically assess the state of the art. We identified and defined five problem areas in this domain: LAN Mistrust, Environment Mistrust, App Over-privilege, No/Weak Authentication and Implementation Flaws. Our assessment led to five security and privacy observations which we utilize to suggest future research directions and expected security practices for building IoT systems. Observations include the lack of substantial studies on emerging IoT application platforms, lack of security solutions for connected cars and the need for device-independent access control solutions to tackle local adversaries. We further highlighted the need for academics, industry representatives and legislators to work together towards a security framework for consumer IoT devices. To facilitate future research, we plan to make available online all our collected data (IoT security articles and IoT apps) and statistics.





%
\bibliographystyle{IEEEtran}
\bibliography{bibliography}

\begin{appendices}
\section{IoT App Security in the Wild}\label{sec:app_analysis}
The most prevalent UIP in the consumer IoT ecosystem is the smartphone. IoT devices are usually controlled by mobile applications, developed either by the devices' respective vendors or other third parties. Our assessment of attack incidents reveal that a notable number of attacks are possible due to hardcoded credentials~\cite{impl2}, lack of encryption~\cite{lack_encryption} and insecure SSL implementations\cite{trusthan3} among others. If a device is vulnerable due to any of these, this would be visible in their mobile application implementations as well. Therefore, instead of limiting our focus on published vulnerabilities, we go a step further and conduct our own real-world study to gain a complete understanding of IoT security. In particular, we collect all apps we could find that connect to IoT devices and perform a security analysis. Although mobile app analysis is not a new topic, there is no previous work focusing on the security of IoT apps. Note that apps might be developed either for iOS or Android. Due to the existence of well-established Android application analysis techniques, we focus on the Android IoT apps.



\vspace {5pt}\noindent
\textbf{App Collection.} Collecting apps that connect to IoT devices is non-trivial. For example, there is no easy way to automatically discover and distinguish such apps on the official Android app store (Google Play) or other third-party markets. In our work, we utilize \url{smarthomedb.com}~\cite{smarthomedb}: smarthomedb retains a mapping between smart home devices and links to their corresponding official mobile apps. We leverage this to build a crawler to automatically collect these links and download the \texttt{apk} files from Google Play. This results in 236 unique apps. To increase coverage, one could leverage Google Play's search engine with relevant keywords. However, such an approach results in a list of apps most of which are irrelevant. In contrast, we utilize another IoT integration platform (\url{iotlist.co}~\cite{iotlist}). We build a tool to automatically extract each listed product's \textit{title} which then use it as the search term in Google Play. Then, the tool retrieves the first result of the query and downloads the corresponding apk file. Lastly, it discards apps that do not fall within the \texttt{House \& Home}, \texttt{Health \& Fitness}, or \texttt{Lifestyle} Google Play categories, which IoT apps usually reside in. This results in 87 more unique IoT apps.

Our approach yields 323 unique IoT apps in total. To validate our tool's effectiveness in retrieving relevant apps, we randomly select 50 discovered apps. We then manually check their descriptions on Google Play. We find that all of them do indeed control IoT devices.





\vspace {5pt}\noindent
\textbf{App Analysis.} Our analysis concentrates on communication security and data protection. Towards this end, we build a static analysis tool based on Androguard APIs~\cite{androguard} which combines several reputable techniques~\cite{conf/ccs/EgeleBFK13, conf/ccs/FahlHMSBF12, conf/blackhat-eu/Yu15}. Our tool totals 1,700 lines of Python code. The tool checks for security issues with respect to three main areas: SSL (S) implementation; Bluetooth (B); Cryptography (C). In particular, the checks are implemented as 16 distinct policies which we summarize below:



\begin{itemize}
	\item \emph{Policy-S1}: App does not use any SSL connection.
	\item \emph{Policy-S2}: Exist URLs that are not SSL protected.
	\item \emph{Policy-S3}: Exist custom SSL \texttt{HostnameVerifier}.
	\item \emph{Policy-S4}: Exist SSL \texttt{AllowAllHostnameVerifier}.
	\item \emph{Policy-S5}: Exist custom SSL \texttt{TrustManager}.
	\item \emph{Policy-S6}: Exist insecure \texttt{SSLSocketFactory}.
	\item \emph{Policy-S7}: Exist custom SSL (WebView) error handlers.
	\item \emph{Policy-B1}: Exist insecure Bluetooth RFCOMM sockets.
	\item \emph{Policy-C1}: App does not use any crypto library.
	\item \emph{Policy-C2}: Exist suspicious Base64 ``encrypted'' strings.
	\item \emph{Policy-C3}: Exist constant / insecure PRNG seeds.
	\item \emph{Policy-C4}: Exist insecure encryption modes (like ECB).
	\item \emph{Policy-C5}: Exist non-random IV for CBC encryption.
	\item \emph{Policy-C6}: Exist constant encryption keys.
	\item \emph{Policy-C7}: Exist constant salts for PBE.
	\item \emph{Policy-C8}: Exist insufficient iterations for PBE.
\end{itemize}

To avoid the interference of irrelevant code, we use a blacklist of more than 100 popular third-party libraries that are not related to network connection or cryptography functions. Most of these are ad libraries, analytics libraries, or user interface libraries.

\begin{figure}[t]
	\centering
	\includegraphics[width=1\columnwidth]{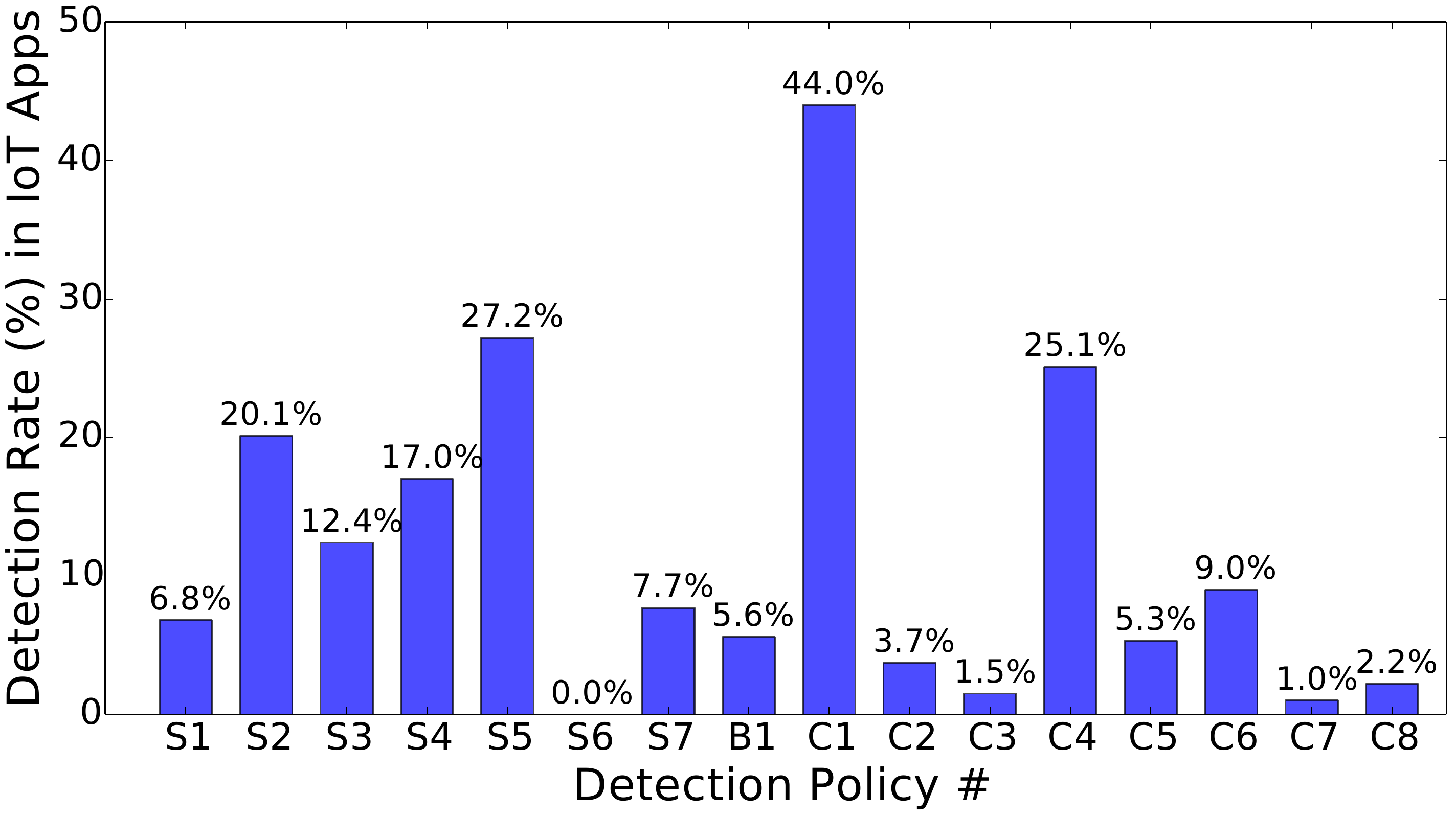}
	\caption{Detection rate of each policy in IoT apps}
	\label{fig:app_detection_rate}
\end{figure}

\vspace {5pt}\noindent
\textbf{Results and Findings.} Our analysis results are alarming as illustrated in Figure~\ref{fig:app_detection_rate}. Nearly 90\% of the IoT apps trigger at least one policy in our detection. In particular, approximately 44\% of the apps do not invoke any crypto function to protect their data (Policy-C1). Among the apps that do use encryption, almost 60\% have insecure implementations of encryption ciphers (Policy-C2$\sim$S8), and 9\% use hardcoded encryption keys (Policy-C6). The situation is much worse regarding SSL implementations. About half of the apps have at least one SSL related flaw (Policy-S2$\sim$S7) or do not use SSL at all for their communications (Policy-S1).


Even though these issues do not necessarily mean that the end IoT devices are exploitable, they do at the very least illustrate that security is in most cases cursively implemented if it is at all. We believe that a unified IoT framework involving industry security standards could help alleviate those issues and guide the development process for IoT apps and device firmwares. To facilitate future studies on IoT, we will release all collected apps online~\cite{iot_sok}.

\begin{mybox}[boxsep=0pt,
	boxrule=1pt,
	left=4pt,
	right=4pt,
	top=4pt,
	bottom=4pt,
	]
~Most of the IoT apps have security implementation issues. This suggests either a lack of security expertise in IoT vendors or the absence of security design and testing in the development process.
\end{mybox}

\end{appendices}
\end{document}